\documentclass[twocolumn, trackchanges]{aastex63}

\usepackage{amsmath}
\newcommand{\Rstorm}{$R_{\rm storm}$}
\newcommand{\smax}{$s_{\rm max}$}
\newcommand{\Ssub}{$S_{\rm sub}$}
\newcommand{\Smass}{$S_{\rm mass}$}

\renewcommand{\v}[1]{{\boldsymbol{#1}}}
\newcommand{\del}{\v{\nabla}}
\newcommand{\grad}{\del}
\newcommand{\Div}{\del\cdot}

\newcommand{\aderiv}[1]{\frac{D{#1}}{Dt}}
\newcommand{\ptderiv}[1]{\frac{\partial{#1}}{\partial{t}}}
\newcommand{\hatx}{\hat{\v{x}}}
\newcommand{\haty}{\hat{\v{y}}}
\newcommand{\hatz}{\hat{\v{z}}}

\usepackage[pagewise]{lineno}
\shorttitle{Exploring the Low Deformation Lengths of Jupiter's Poles}
\shortauthors{Hyder et al.}
\graphicspath{{./}{figures/}}
\begin{document}

\title{Exploring Jupiter's Polar Deformation Lengths with High Resolution Shallow Water Modeling}

\author{Ali Hyder}
\affiliation{Department of Astronomy, New Mexico State University, PO BOX 30001, MSC 4500, Las Cruces, NM 88003-8001, USA.}

\author{Wladimir Lyra}
\affiliation{Department of Astronomy, New Mexico State University, PO BOX 30001, MSC 4500, Las Cruces, NM 88003-8001, USA.}

\author{Nancy Chanover}
\affiliation{Department of Astronomy, New Mexico State University, PO BOX 30001, MSC 4500, Las Cruces, NM 88003-8001, USA.}

\author{Ra\'ul Morales-Juber\'ias}
\affiliation{Department of Physics, New Mexico Institute of Mining and Technology, Socorro, NM 87801, USA.}

\author{Jason Jackiewicz}
\affiliation{Department of Astronomy, New Mexico State University, PO BOX 30001, MSC 4500, Las Cruces, NM 88003-8001, USA.}

\begin{abstract}

The polar regions of Jupiter host a myriad of dynamically interesting phenomena including vortex configurations, folded-filamentary regions (FFRs), and chaotic flows.  \textit{Juno} observations have provided unprecedented views of the high latitudes, allowing for more constraints to be placed upon the troposphere and the overall atmospheric energy cycle.  Moist convective events are believed to be the primary drivers of energetic storm behavior as observed on the planet.  Here, we introduce a novel single layer shallow water model to investigate the effects of polar moist convective events at high resolution, the presence of dynamical instabilities over long timescales, and the emergence of FFRs at high latitudes.  We use a flexible, highly parallelizable, finite-difference hydrodynamic code to explore the parameter space set up by previous models.  We study the long term effects of deformation length ($L_d$), injected pulse size, and injected geopotential.  We find that models with $L_d$ beyond $1500$ km (planetary Burger number, Bu$=4.4\times10^{-4}$) tend to homogenize their potential vorticity (PV) in the form of dominant stable polar cyclones, while lower $L_d$ cases tend to show less stability with regards to Arnol'd-type flows.  We also find that large turbulent forcing scales consistently lead to the formation of high latitude FFRs.  Our findings support the idea that moist convection, occurring at high latitudes, may be sufficient to produce the dynamical variety seen at the Jovian poles.  Additionally, derived values of localized horizontal shear and $L_d$ may constrain FFR formation and evolution.

\end{abstract}


\defcitealias{Brueshaber+2019}{BSD19}

% original intro at line 4
\section{Introduction} \label{sec:intro}

The variety of cloud morphologies seen in Jupiter's visible troposphere is extensive, making it a rich laboratory for the study of a wide range of dynamical processes.  It contains planetary-scale wind fields that are comprised of high speed jets between the belts and zones, along with long-lived storm features such as the Great Red Spot (GRS) at the lower latitudes.  Although Jupiter's jet streams extend up to the mid-latitudes, there is a distinct difference in appearance beyond $\pm60^{\circ}$ \citep{Orton+2017}, where there is a complete disintegration of the horizontal banded structures of the lower to mid-latitudes, and the local turbulent flows self-organize into more discretized vortex behavior.

An important discovery made by the Jovian InfraRed Auroral Mapper (JIRAM) \citep{jiram} instrument on board the \textit{Juno} spacecraft was that of clustered polar cyclones at the high latitudes, also referred to as vortex configurations \citep{Adriani+2018}.  Such behavior is in stark contrast to Saturn's polar hexagon \citep{Godfrey1998} and tends to emerge in 2D Euler flows.  \cite{JinDubin1998} showed the emergence of such features in idealized cases using regional maximum entropy theory, which states that vortex configurations form due to strong vortices mixing the surrounding fluid, maximizing the local entropy, thereby limiting the underlying chaotic motion and stabilizing the system.  However, the mechanisms through which such structures emerge in an ever-changing atmosphere are unclear.  It is important to understand what mechanisms are involved in the formation, structure, and stability of cyclonic configurations to better constrain the overall atmospheric energy cycle of Jupiter.

Recently, 3D polar simulations were used to model the formation of such configurations via turbulent convection, exploring the role of inertial stability in a convectively unstable background \citep{Cai2021DeepCP}.  Although this model successfully showed the emergence of 3D polar configurations, here we show how a simplified 1.5-layer shallow water model can also capture relevant dynamics pertinent to such instabilities.  Such a model allows for an exploration of a larger parameter space due to lower computational demands.

For the case of Jupiter, a host of simulations have been performed to understand the global circulation of the system and characterize the energy cycle.  This includes, but is not limited to, full general circulation models of the entire atmosphere \citep[e.g.][]{Williams1978, Williams1988, Kaspi+2009}, as well as localized process modeling for predictions of storm behavior and evolution to determine the quasi-steady state of such systems \citep[e.g.][]{GARCIAMELENDO2017}.  Shallow water models of Jupiter's weather layer have provided significant insights on the nature of the Jovian dynamics, including the emergence of super rotating jets and long-lived storms \citep{Dowling1993, Showman_2007}.  However, models that focus on the high latitudes, specifically the polar regions, have been limited due to geometric constrains and low resolution.

\cite{ONeil+2015} performed a comprehensive analysis of gas giant polar regions using a 2-layer shallow water system forced by `hetons', which are defined as 2-layer vortices with opposite rotation in the vertical direction \citep{hetons_1985, Pedlosky1085, YANO1992}.  They found that the planetary Burger number, defined as
\begin{equation}
{\rm Bu} = (L_d / a)^2 ,
\label{e:Bu}
\end{equation}

\noindent quantified the overall equilibrated dynamics and morphology of the atmosphere.  $L_d$ is defined as the deformation length and $a$ is the planet's radius of curvature at the pole.  It is important to note that the planetary Burger number is distinct from the \textit{local} Burger number, Bu$_{l} = (L_d/R_{\rm storm})^2$, where $R_{\rm storm}$ is a storm radius.  Here, we use the term ``morphology" as a qualitative measure of what is readily observed in the models, specifically after equilibration of the storm forcing in the long-term.

\citet[hereafter \citetalias{Brueshaber+2019}]{Brueshaber+2019} used the 1.5-layer shallow-water capability of the Explicit Planetary Isentropic Coordinate General Circulation Model \citep[EPIC GCM;][]{Dowling+1998} to investigate mechanisms involved in such morphological differences across the upper atmospheres of the giant planets.  They quantified the effects of cyclone to anticyclone ratio, storm strength, and planetary Burger number, Bu, on the long-term evolution of their models, and also found Bu to be the most important dimensionless parameter that characterizes the primary mode of the dynamics.  \citet{Brueshaber+2021} further explored Bu values relevant for Saturn and Saturn-like systems to study the detailed behavior of emergent large-scale polar cyclones that dominate most of the polar region.

In this study, we build upon previous work by developing a new shallow water code that circumvents the polar singularity using the polar-$\beta$ plane approximation, also called the $\gamma$-plane.  The model injects storms stochastically in the domain, resulting in 2D turbulence and long-term dynamics.  Using our model, we attempt to reproduce the range of observed Jovian polar morphologies using previously employed parameterizations of moist convection \citep{ONeil+2015, Showman_2007}.  In \S \ref{sec:modeling}, we provide non-dimensionalization of the main parameters and details of our model.  In \S \ref{sec:results}, we present results from our model.  In \S \ref{sec:furtherdisc} we discuss our results regarding the emergence of folded-filamentary behavior and low deformation length effects, and in \S \ref{sec:conclusion} we present our conclusions and plans for future work using this model.  We provide scalability tests and benchmarks that support the validity of the model in Appendix \ref{sec:app_scaling} and \ref{sec:app_Benchmarks}, respectively.

\section{Model Description} \label{sec:modeling}

Modeling efforts in the Jovian atmosphere community employ well-tested and robust codes.  These include (but are not limited to) the EPIC GCM, Isca \citep{Isca_2018}, DYNAMICO \citep{DYNAMICO_a, DYNAMICO_b}, and the MITgcm \citep{MITgcm}.  These codes provide a multitude of numerical tools necessary to model phenomenologically distinct scales of gas giant atmospheres.  For the case of Jupiter, EPIC has served as the workhorse for those interested in the full primitive equations, while the MITgcm has been used extensively for the anelastic approximation \citep{Kaspi2008PhD}.  Here, we introduce an atmospheric module for the shallow water approximation using the {\sc Pencil Code}\footnote{{\sc{The Pencil Code}} is available for download at {\url{http://pencil-code.nordita.org/}}}.

Pencil is a $6^{th}$-order spatial and $3^{rd}$-order temporal finite-difference code, well-suited for modeling three dimensional weakly compressible turbulence \citep{Brandenburg+2002, BrandenburgDobler10, Brandenburg+Scannapieco2020}.  It uses a collocated Eulerian mesh instead of a staggered-C grid that is widely used in other codes.  Staggered grids offer stability with second-order spatial accuracy.  Pencil, on the other hand, offers higher order spatial accuracy using a simple collocated grid with explicit centered finite differences.  Although that is not suitable for second-order schemes, Pencil achieves high accuracy due to the high-order nature of the stencil.  It has been heavily employed in models of the solar convective turbulence in the solar corona \citep[e.g.][]{Bourdin+2013}, hydrodynamical instabilities in protoplanetary disks \citep[e.g.][]{Raettig+13, Raettig+2015, Lyra+15, Lyra+17, Lyra+18}, and is suitable for weakly compressible hydrodynamical and hydromagnetic turbulence \citep{Haugen+2004}.  Here, we present the first application of Pencil to atmospheric dynamics with a focus on the meteorological processes of Jupiter's polar region.  Although Pencil has the capability to perform 3D numerical simulations, we limit our work to the shallow water approximation to make use of the high resolution that can be afforded to 2D models.

In order to accurately simulate the non-linear fluid dynamics, the first-baroclinic Rossby radius of deformation must be resolved. For polar models, typical values for the Rossby radius of deformation of around $2000$ km have been used, even though deformation lengths of that scale correspond to equatorial and mid-latitude regions (although \citetalias{Brueshaber+2019} used $1000$ km along with a significantly large planetary radius to simulate a low Bu environment).  The highest resolution achieved in their model was $512^2$ ($\Delta x \approx 150$ km), which is sufficient for stable evolution but is not enough to resolve detailed behavior of the large cyclonic and anticyclonic regions with increasingly low deformation length.  Previous works have generally explored the poles of Jupiter and Saturn with a resolution of $256^2 - 512^2$ \cite[e.g.][]{Morales-Juberias+2011, Morales-Juberias+2015, ONeil+2015, Brueshaber+2021}; for Jupiter size scales, such resolutions correspond to $\Delta x = 150 - 300 {\rm\ km}$.  Our model takes the base resolution further by exploring Jupiter's dynamics with a resolution of $1024^2\ (\Delta x = 75 {\rm\ km}, 4 \times$ more resolution in area than previous shallow water models$)$, which proves necessary to study the long term effects of the lowest deformation lengths and resolved forcing scales that might be dynamically relevant for atmospheric evolution.  We test the model up to a resolution of $2048^2$ ($\Delta x = 38 {\rm\ km}, 16 \times$ more resolution than earlier models).  However, as we explore a vast parameter space, we limit model resolution to $1024^2$.

\subsection{Shallow Water Approximation}

Although multiple layers can be used, we limit our analysis to a single-layer forced turbulence approach, thereby restricting the parameter space to barotropic instabilities.  Perturbations representing moist convection are made directly to this layer.  The bottom layer, often referred to as the ``$1/2$-layer", represents the deep atmosphere, which is assumed to be quiescent.  Such systems are conventionally called 1.5-layer models.  In multilayer systems, these perturbations can be generated deeper and their effects understood in context of the reaction of the surrounding layers.  Single layer systems are limited in terms of the overall dynamics and energy equilibration as they do not permit baroclinic modes.  Although restricted, they are sufficient for studying the long term evolution of non-stratified systems.  Since the system is limited to a single layer with a ``quiescent" deep layer, the momentum equations in the corotating frame reduce to the following:
\begin{eqnarray}
\label{eq:masscont}
\frac{D (gh)}{Dt} &=& - gh \Div{\v{u}} +\sum^{n}_{i=1} S_{i, \rm storm} + S_{\rm sub} + S_{\rm mass} \nonumber \\
&& + D_3 \nabla^6 (gh) + \Div{\v{J}},\\
\label{eq:momcont}
\frac{D \v{u}}{Dt} &=& -\grad{\left(gh\right)} -f\hatz\times\v{u} + \nu_3 \nabla^6 \v{u} + \Div{\v{\xi}}
\end{eqnarray}

\noindent where $g$ is the reduced gravity, $h$ is the fluid height ($gh$ is the ``geopotential"), $f$ is the Coriolis parameter, and $\hatz$ is the unit vector in the vertical (outward) direction.
\begin{equation}
  \aderiv{} \equiv \ptderiv{} + \v{u}\cdot\del    
\end{equation}

\noindent is the Lagrangian derivative operator, and $\v{u} = u\hatx + v\haty$ is the horizontal velocity. The coefficients $D_3$ and $\nu_3$ are  $6^{th}$-order hyperdiffusion and hyperviscosity coefficients, respectively, needed for numerical stability such that dissipation occurs with a higher-order differential rather than the standard Laplacian form, maximizing the inertial range. The last terms in the mass and momentum equations represent divergence damping, also called divergence limiting \citep{Skamarock&Klemp1992}, and take the form of mass diffusion and a simplified bulk viscosity
\begin{eqnarray}
  \v{J} \equiv D_{\rm bulk} \grad{(gh)},\\
  \label{eq:bulk-diff}
  \xi_{ij} \equiv \nu_{\rm bulk} \, \partial_j u_i.
  \label{eq:bulk-viscosity}
\end{eqnarray}

Divergence damping is employed to filter out high frequency acoustic waves that are not relevant for long timescale atmospheric dynamics \citep{Klemp+2018}.  The coefficient $\nu_{\rm bulk}$ is an artificial bulk viscosity.  We also define $D_{\rm bulk}$ as an artificial ``bulk diffusion". They are equal in value, defined numerically as

\begin{equation}
  D_{\rm bulk } = \nu_{\rm bulk } \equiv c_{\rm bulk}  \left<\max_5[(-\Div\v{u})^+]\right>{\left[\min(\Delta x)\right]}^2{.}
  \label{eq:bulk-visc-coeff}
\end{equation}

\noindent where the superscript plus sign indicates the positive part of the quantity. This formulation requires that the divergence limiters are proportional to the maximum of positive flow convergence, as evaluated over five grid cells in each direction for a total of 25 zones in 2D  (the given cell plus its immediate neighbors). The angled brackets represent a quadratic smoothing function that smooths the divergence over seven zones in each direction, with weights (1, 6, 15, 20, 15, 6, 1)/64. The result is then scaled by the square of the smallest grid spacing. The quantity $c_{\rm bulk}$ is a constant defining the strength of the bulk diffusivities, set to unity in our simulations. The divergence limiters in the equations of motion have the effect of smoothing a numerical discontinuity until it is resolved by the stencil. Notably, the formulation in terms of divergences of diffusion fluxes implies that these extra terms, although artificial, conserve mass and momentum. This type of divergence limiter, taking the mathematical form of a bulk viscosity, is routinely used in compressible hydrodynamical simulations to treat shocks \citep[e.g.][]{Haugen++2004, Richert+15, Lyra+16, Hord+17}.

$S_{i, \rm storm}$ represents the forcing model and drives the geopotential gradients via stochastic perturbations (iterated over $i$) to which the velocity fields must adjust geostrophically.  The sum is conducted over each perturbation.  The total number of active injected storms remains fixed throughout this work and is maintained at $n = 38$.  This is sufficiently high for large scale dynamics and interactions to occur over long timescales but is low enough that emergent features are not immediately disrupted by concurrent moist convective events.  The effect of varying $\textit{n}$ is not studied in this work. The choice of  $\textit{n}$ is motivated by scaling the $100$ storm injections used by \citetalias{Brueshaber+2019} for a planetary radius of $200,000$ km.  As we use Jupiter's radius explicitly, we lowered the number of active storms from 100 to 38.  \Ssub~adds or removes the overall mass outside of an injection site to ensure that the domain-averaged geopotential and deformation length does not vary \citep{ONeil+2015, ONeill+2016}.  Finally, \Smass~represents the mass relaxation that removes mass by steadily bringing the instantaneous average thickness to the steady-state over the relaxation timescale.  The forms of equations \ref{eq:masscont} and \ref{eq:momcont} closely follow the formulations provided in BSD19 and \cite{Brueshaber+2021} as the fundamental equations of motion for both models are the same, apart from the $6^{th}$-order hyperviscosity term that is implemented in our case.

We provide perturbations to the system as Gaussians in space and time following BSD19:
\begin{equation}
S_{\rm storm} \equiv s_{\rm max} \exp{\left( -\frac{r^2}{R_{\rm storm}^2} - \frac{(t-t_{\rm p})^2}{\tau^2} \right)},
\label{e:Sstorm}
\end{equation}

\noindent where \smax\ is the maximal injection rate, $r$ is the mass injection site offset from the pole, \Rstorm\ is the peak injection pulse size (or perturbation radius), and $\tau$ is the characteristic mass injection time, which is set to $10^5\,$s.  $t_{\rm p}$ is the peak mass injection time, which is set to half of the storm duration.  The interval between storm injection is set to $2\tau$.  The mass relaxation term, $S_{mass}$, follows the same moist convection formulation as in \citet{Showman_2007}.  The subsidence term, $S_{sub}$, adds or removes the mass injected from a given vortex over the whole domain, ensuring that geopotential perturbations do not cause a variation in the steady-state geopotential \citepalias{Brueshaber+2019}.

Typically, the shallow water continuity equation is written solely in terms of the fluid height, $h$.  However, since the reduced gravity, $g$, remains constant for the 1.5-layer model, we analyze our simulations in terms of the geopotential, $gh$ \citep{Dowling&Ingersoll1989}.  We employ a forced turbulence model in our system, where the perturbations only occur on the geopotential \citep{ONeil+2015}.  This allows for horizontal velocities to respond to perturbations in the flow and stabilize accordingly.

The mass injections are parameterizations of moist convective events, the details of which cannot be incorporated in a shallow water approximation.  Our model captures the effective resultant behavior due to such storms being churned up, perturbing the dynamics, and the consequent winds adjusting geostrophically.  Such parameterization captures the bulk properties of mass and energy perturbations due to moist convection, even though complexities of the process are ignored.  The perturbation imparts surface gravity waves that dissipate some of the energy away but most of it remains at the injection site, driving winds whose maximum velocities depend upon the storm intensity applied.  We thus use estimates of the wind speeds from observations \citep{Orton+2017, Vakili+2020} to sufficiently constrain the injection strength to ensure we are not overforcing.  The injection strengths that produce realistic wind speeds are used as upper limits of the forcing model.

\subsection{Non-dimensionalization of Physical Parameters}

Non-dimensionalization can reduce the overall complexity of the problem and provide insight into the key length and time scales of a system.  In order to properly reproduce the polar dynamics in an appropriately scaled manner, we must ensure that we use the correct non-dimensional values of the parameters we are interested in.  We begin our non-dimensionalization with time.  A characteristic timescale for the dynamics of Jupiter can be obtained via its rotation frequency, $\Omega_J = 1.76\times10^{-4}\,$s$^{-1}$.  However, since we are using the Rossby deformation radius as a given known parameter, we may also determine the Coriolis frequency, $f_0 = 2\Omega$, using the deformation length,
\begin{equation}
L_d \equiv \sqrt{\langle gh \rangle} / f_0,
\end{equation}
where $\langle gh \rangle$ is the steady-state geopotential.  The unit of time, $[t]$, is defined as,
\begin{equation}
[t] \equiv \frac{1}{\Omega} = \frac{2}{f_0} = \frac{2 L_d}{\sqrt{\langle gh \rangle}}
\end{equation}

The choice of steady-state geopotential, $\langle gh \rangle$, and deformation length sets the rotation rate for the planet.  Here, the angled brackets around the geopotential denote a domain-average.  This results in two distinct methods for choosing the rotation: using the observed rotation rate of Jupiter, thereby setting $\langle gh \rangle$ uniquely for a given model, or by varying the rotation rate of the planet explicitly.  The two approaches are equivalent in their non-dimensional form but allow for a distinction between the relevant parameters being tested.  The bottom layer of the fluid can have spatial variations across the domain, often referred to as `dynamical topography' \citep{Dowling2020}.  For the purposes of this study, we do not implement any such spatial variations as that would significantly increase the parameter space we explore in this work.

The most appropriate unit of length for these polar simulations is the Jovian osculating radius, $a = (R_{e}^2/R_{p}) = 76,452$ km, where $R_e$ and $R_p$ are Jupiter's equatorial and polar radii, respectively\footnote{$R_e = 71,492$ km; $R_p = 66,854$ km \citep{jupiter2007}.}.  Thus, varying the deformation length directly changes the planetary Burger number, which is a function of planetary radius and the Rossby deformation length.  For each corresponding planetary Bu value, we produce models with mass injection sizes that satisfy \Rstorm$/L_d < 1$ (\textit{local} Burger number, Bu$_{l} > 1$), effectively exploring a storm-to-deformation length ratio.  The deformation radius represents the length scale where the pressure gradient of a storm, acting outwards (inwards), is balanced by the Coriolis force acting inward (outward).  At this length scale, rotational effects become as important as local geostrophic behavior.  The deformation radius further modulates the overall distance two vortices can remain from each other without non-linear interactions causing disruption or coalescence.  The planetary Burger number, therefore acts as a proxy for rotational dominance in an effectively non-stratified environment, suitable for the shallow water approximation.

Following \citetalias{Brueshaber+2019}, we set a steady-state geopotential, $\langle gh \rangle = 2\times10^{5}$ m$^2$\,s$^{-2}$.  This parameterization represents a model suite with varying differential rotation, ranging from $0.63\,\Omega_J$ to $1.69\,\Omega_J$.  Since we maintain a fixed domain-averaged geopotential through a given suite of simulations, the timescale solely depends on the value of the planetary Burger number via equation \ref{e:Bu}, with smaller Bu implying faster rotation, as ${\rm Bu} \propto L_d^2 \propto t^2$.  The corresponding non-dimensionalizations relevant to our model suite are presented in Table \ref{table:nondim}.  For the timescale, we present both non-dimensional scalings in the code depending on the choice of fixed geopotential or fixed rotation.  Finally, as we extend the application of previous work, we define the nondimensional eddy potential vorticity as in \citetalias{Brueshaber+2019} to facilitate an easier comparison:
\begin{equation}
Q_e^* = \left[ \frac{\zeta + f}{h} - \frac{f}{\langle h\rangle} \right] \cdot \frac{\langle h\rangle}{f_0},
\label{e:Q_e}
\end{equation}

\noindent where $\zeta$ is the local relative vorticity, $\langle h\rangle$ is the steady-state fluid thickness, and $f_0 = 2\Omega$.

\begin{table}[ht]
\caption{Nondimensionalization of relevant parameters.  These are the scales used to convert from the code units used herein to physically meaningful values.}
 % title of Table
\centering % used for centering table
\begin{tabular}{c c c} % centered columns (4 columns)
\hline\hline %inserts double horizontal lines
Physical parameter & Model conversion factor \\ [0.5ex] 
\hline % inserts single horizontal line
Length & $a$ \\
Time & $2L_d / \sqrt{\langle gh \rangle}$ or $1/\Omega_{J}$ \\
Energy & $a^6 \Omega^4$ \\ 
Mass injection & $a^2 \Omega^3$ \\ [0.5ex] % [1ex] adds vertical space
\hline %inserts single line
\end{tabular}
\label{table:nondim} % is used to refer this table in the text
\end{table}

\subsection{Geometry and the Gamma-Plane Approximation}

\begin{figure*}[hbtp] \center
    \includegraphics[width=4.25in, height=3in]{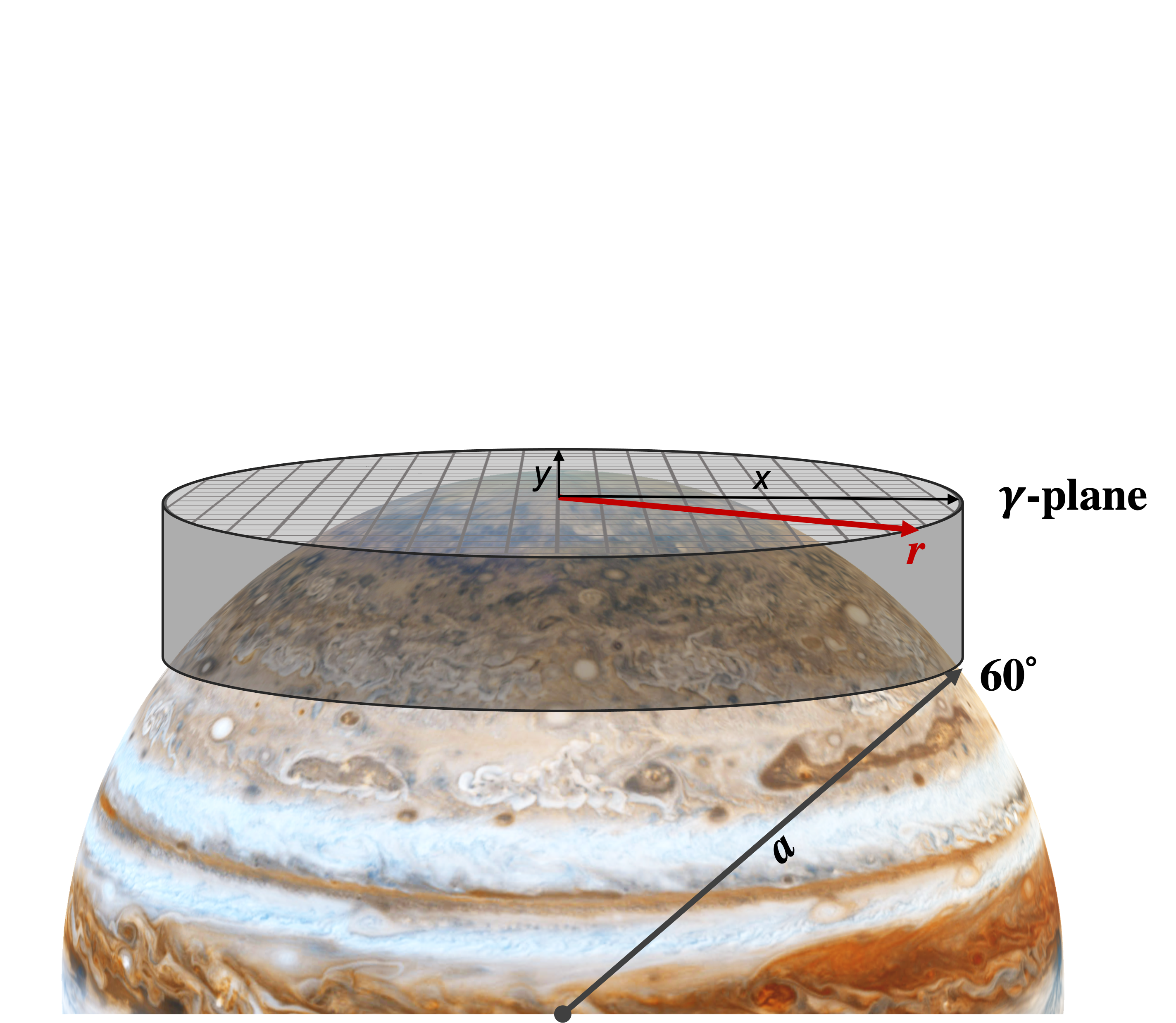}
    \caption{Illustration of the $\gamma$-plane.  The approximation is equivalent to a $\beta$-plane at the lower latitudes but applied to the pole.  The polar coordinates are converted to a Cartesian grid for the numerical computation to avoid the polar singularity.}
    \label{f:gammaplane}
\end{figure*}

We use a Cartesian approximation for the model to circumvent pole-centered singularities that emerge due to a polar geometry.  Although other methods are applicable at the poles, the Cartesian approximation sufficiently captures the main morphological features of the atmosphere in our model.

The $\gamma$-plane approximation, as used in \citetalias{Brueshaber+2019} and also employed in the form of the polar-$\beta$ plane by \cite{ONeil+2015}, allows an exploration of the shear effects dominating the high latitudes.  It quantifies the planetary vorticity gradient and offers parameterization of the length and time scales directly.  In short, $\gamma$ carries the effect of variations in the planetary Burger number.  With a fixed planetary radius, it traces the behavior of the deformation length.  The functionality is implemented using the following correction to the Coriolis parameter at the pole:
\begin{equation}
f = 2\Omega - \gamma r^2,
\end{equation}

\noindent where $\gamma = \Omega/a^2$ \citep[\citetalias{Brueshaber+2019}]{Bridger&Stevens1980}.  Our model uses the Jovian osculating radius, $a$, as the domain size and use a doubly periodic domain.  We model latitudes poleward of $60^{\circ}$ North.  Figure \ref{f:gammaplane} illustrates the setup of the $\gamma$-plane with the natural polar coordinates, $(r, \theta)$ where $\theta$ is the azimuthal angle, converted to Cartesian coordinates $(x, y)$ for our model.  The domain of interest extends from the center of the box to a circle of $0.5a$ in radius, after which a damping parameter is applied such that a traversing gravity wave is quickly damped outside the domain.  Strongest damping occurs farthest away from the outer circle to minimize shear generation due to the boundary.  We apply this damping to a variable $X$ in the form of a relaxation term over a timescale $\tau_{\rm dis}$, so that
\begin{equation} \label{eq:dragterm}
\frac{\partial X}{\partial t} = -\frac{X - X_0}{\tau_{\rm dis}}\sc{S(r)},
\end{equation}

\noindent where $X_0$ is the desired value.  $\sc{S(r)}$ is a $5^{th}$-order step function that smoothly goes from zero to unity from $0.5a$ to $0.55a$, respectively.  The timescale $\tau_{\rm dis}$ is chosen such that a gravity wave dissipates within this buffer zone.  Beyond $0.55a$, a freezing boundary condition is used where all derivatives are set to zero \citep{Lyra+08a}.  Thus, similar to \citetalias{Brueshaber+2019}, no vortex feature or propagating gravity wave can go across the damping region and reenter the domain.

\section{Results} \label{sec:results}

The Jovian polar cyclones have an approximate range of sizes from $4000 - 7000$ km and exhibit velocities between $\sim 55 - 115$ m/s as measured from JunoCam \citep{Vakili+2020} and JIRAM \citep{Grassi+2018} images.  We restrict our model to produce wind speeds comparable to observed velocities, providing a way to constrain the turbulent forcing strength necessary to sufficiently perturb the steady-state geopotential.  Given $\langle gh \rangle = 2\times10^{5}$ m$^2$\,s$^{-2}$, we used mass injections that would produce wind speeds on the order of 100 m/s.  Previous models (e.g., \cite{ONeil+2015}; \citetalias{Brueshaber+2019}; \cite{Brueshaber+2021}) used a range of perturbations but exact values of injection strengths are not well-constrained.  Our forced turbulence model has two forms for two distinct applications: varying Burger numbers (planetary and local) and varying storm strengths, described in \S 3.1.1 and 3.1.2 as Case A and Case B, respectively.  For the case that utilizes previously published values for a steady-state geopotential (Case A), we use $s_{max}=0.25$ m$^2$\,s$^{-3}$, which proves sufficient for our model as the resulting wind speeds are in agreement with derived velocities.  For the case of Jupiter's rotation (Case B), we vary the storm strength such that total \textit{injected} geopotential is reasonably small compared to the fluid thickness for that simulation.  Otherwise, the same perturbation strength for a lower geopotential will result in overforcing.  This method defines a unique geopotential injection parameter that allows us to explore the effect of forcing strength in an atmosphere with Jupiter's rotation.  The details of the injection strength are provided in \S \ref{ssec:CaseB}.

\subsection{Parameter Space}

As there are two methods for choosing the timescale for the models, we explore both in order to study the different parameters.  Previous models of the polar region focused more on determining the general characteristics of gas giant polar dynamics, and thus used timescales derived from a choice of deformation length and steady-state geopotential (as opposed to using Jupiter's differential rotation), as well as significantly larger planetary radii.  In contrast, we used Jupiter's osculating radius, $a$, explicitly as the length scale for all of our models and explore the parameter space by studying both time scales independently.  This results in two cases.  In Case A, we vary the planetary and local Burger number by varying deformation lengths and storm size.  The specified steady-state (domain-averaged) geopotential and choice of deformation length determine the timescale, $\Omega^{-1}$, for these simulations (Table \ref{tab:LdvsRs}).  This implies a unique rotation for each choice of deformation length.  In Case B, we vary the storm strength and use the observed System III rotation rate of Jupiter, thereby varying the underlying steady-state geopotential for a choice of deformation length.  Our simulations produce a variety of morphologically distinct and interesting structures.  These behaviors are a result of the higher resolution and computational accuracy of our model.  Our model shows similarity with previous work, in particular \cite{ONeil+2015}, \citetalias{Brueshaber+2019}, and \cite{Brueshaber+2021}, and additionally shows the emergence of distinct novel behavior as described in detail in the following subsections.

\subsubsection{Case A - Varying Planetary \& Local Burger Numbers}

We use the ratio \Rstorm$/L_d$ as one of the parameters in our models.  Although this ratio is the inverse square root of the local Burger number, Bu$_l$, we vary the ratio itself to have a more direct relation to the storm size.  \cite{Brueshaber+2021} employed this parameter to explore the dynamical regimes relevant to Saturn and Saturn-like systems, i.e. higher planetary Burger numbers.  Here, we provide a completion of the parameter space by studying the effects on Jupiter with lower planetary Burger numbers by using smaller deformation lengths with higher resolution.  The values employed in the model suite are physically motivated due to the overall dynamical effect of the Rossby deformation length.  If Bu$_l < 1$, the forced turbulence model would represent nonphysical storm injections as the imparted pulse would be larger than the deformation length.  As our mass injections represent physically relevant moist convective events, a deep moist convective plume would be undergoing deformation due to the Coriolis force as it emerges in the upper atmosphere.  Consequently, it would not be larger than the overall deformation length.  With Bu$_l > 1$, all emergent storms are such that they grow by coalescence or aggregation of surrounding fluid until they attain a phase of geostrophic adjustment.  This does not preclude the existence of storms with Bu$_l < 1$ as such features may form via mergers, but only limits the type of moist convective plume length scales relevant in the forcing model.

For a given deformation length, the \Rstorm values are tested from $522-1750$ km in increments of $250$ km (except for the first increment).  The lowest \Rstorm value is kept at $522$ km as we maintain a $6^{th}$-order spatial derivative across the radius, fully resolving the injected pulse.  This is a strict constraint that is maintained throughout the study.  The deformation lengths themselves are varied from $750-2000$ km (Bu $\sim 9.6\times10^{-5} - 6.8\times10^{-4}$).  This defines a single simulation for the $L_d=750$ km case, two for $L_d=1000$ km, three for $L_d=1250$ km, four for $L_d=1500$ km, five for $L_d=1750$, and six for $L_d=2000$ km.  The results for this model suite are presented in \S \ref{ssec:DepDefLen}.

\subsubsection{Case B - Varying Storm Strength} \label{ssec:CaseB}

We ran the simulations in Case A using a fixed domain-averaged geopotential that was maintained for the whole parameter space, similar to what was done in \citetalias{Brueshaber+2019} and \citet{Brueshaber+2021}.  However, the consequence of such a system is that the suite represents simulations with varying rotations.  To ensure the applicability of the model to Jupiter's case, we relax this constraint.  Using Jupiter's rotation, $\Omega_J = 1.76\times10^{-4}$ s$^{-1}$, we must vary the storm strengths to prevent over-forcing of the fluid layer.  The time-dependent component of $S_{\rm storm}$ (equation \ref{e:Sstorm}) may be integrated as we do not vary the temporal components of the forcing model.  Integrating over injection duration, $t_{d}=2.2\tau$ {\bf{\citepalias{Brueshaber+2019}}}, the injected storm geopotential, $gh_{\rm inj}$, is given by

\begin{equation}
\begin{split}
gh_{\rm inj} & = s_{\rm max} \int_0^{t_{d}} \exp{\left[- \frac{(t-t_{p})^2}{\tau^2}\right]} dt, \\
         & = s_{\rm max}  \frac{\sqrt{\pi}}{2}\tau\left[{\rm erf}\left(\frac{t_{p}}{\tau}\right) - {\rm erf}\left(\frac{t_{p} - t_{d}}{\tau}\right) \right],
\end{split}
% \label{e:IntegralSstorm}
\end{equation}

\noindent where we have evaluated the spatial component at the injection site ($r=0$).  The values of \smax~ are chosen such that the mass injection never exceeds $\sim20\%$ of the total fluid thickness, which ensures that the produced winds will remain within observed values.  We define a threshold parameter for the forcing using the ratio of the geopotentials as
\begin{equation}
\Phi_r \equiv \frac{gh_{\rm inj}}{\langle gh \rangle} \times 100\%,
\label{eq:Delta}
\end{equation}

\noindent where $\langle gh \rangle$ is the domain-averaged steady-state geopotential given for the choice of $L_d$.  We express it as a percentage of the total geopotential and vary it incrementally to observe the overall effect of storm strength over fixed rotation for the cases of $L_d = (750, 1250, 2000)$ km, producing a suite of $9$ models.  The details of the models are presented in \S \ref{ssec:forcingstrength}.

\subsection{Dependence on Deformation Length} \label{ssec:DepDefLen}

\begin{figure*}[hbtp] \center
    \includegraphics[width=\textwidth]{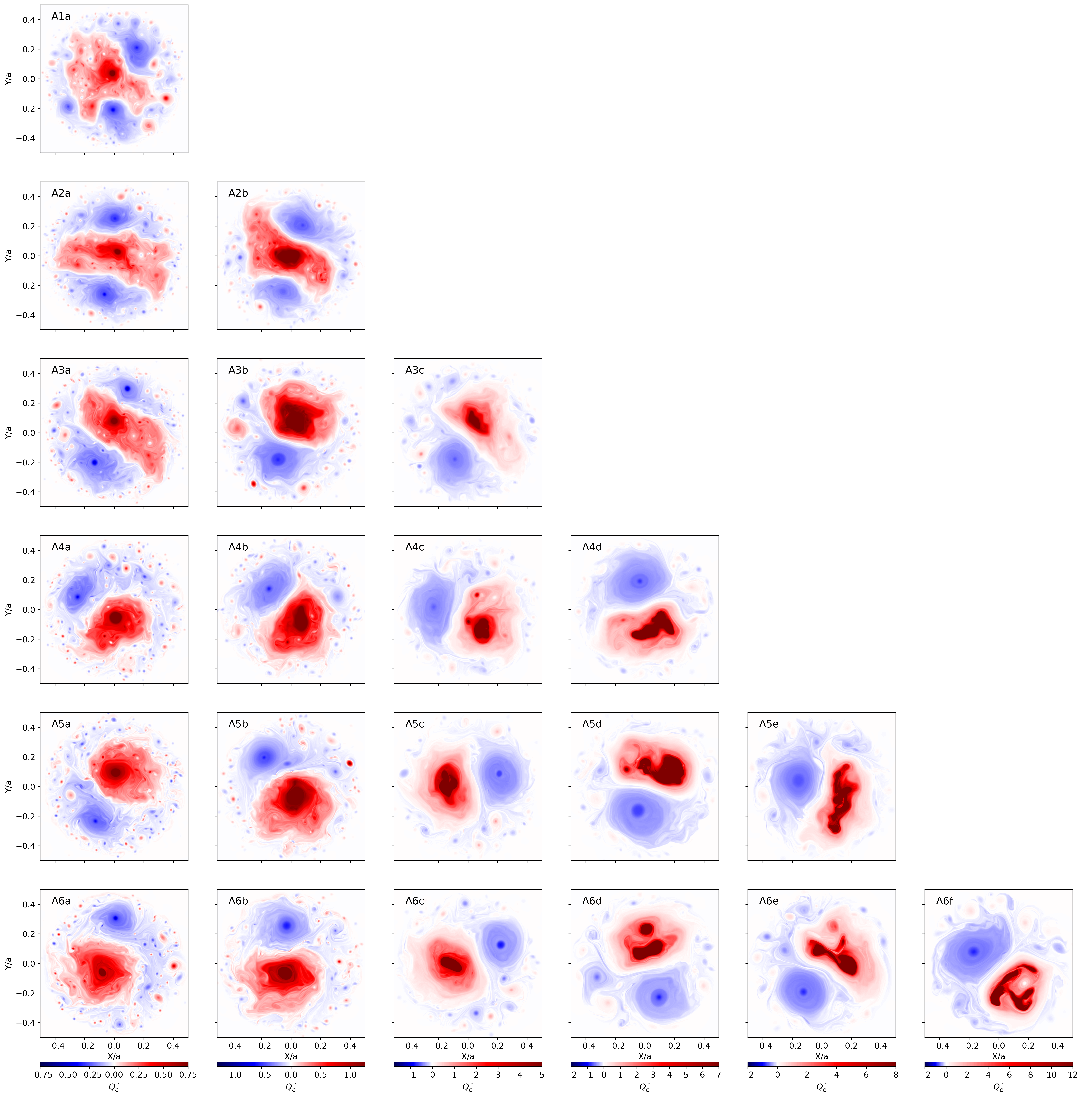}
    \caption{Final snapshots of Case A simulations.  Non-dimensional eddy potential vorticity, $Q_e^*$, as a function of deformation length and injected pulse size at day 20,000.  We use the same color scheme as \citet{ONeil+2015, ONeill+2016} where red indicates cyclonic motion and blue indicates anticyclonic motion.  \textit{Top to bottom:} Increasing deformation length, $L_d$, from $750$ km to $2000$ km with $250$ km increment. \textit{Left to right:} Increasing the injected pulse size, $R_{storm}$, from $\approx 500$ km to $1750$ km with $250$ km increment.  Each colorbar pertains to the column of simulations above it.  Resolution: $1024^2$ ($\Delta x = 75$ km).  Movies for these Case A simulations are available on Zenodo: \dataset[doi:10.5281/zenodo.6642986]{https://doi.org/10.5281/zenodo.6642986} \citep{hyder_ali_2022_6642986}. The movies show that primary distinction between the deformation lengths manifest early on even though the overall dynamics continue to evolve.}
    \label{f:LdvsRs}
\end{figure*}

We produced a suite of 21 simulations that parameterize the entire domain of the observed deformation lengths on Jupiter, and their relation to the size of the injected pulse.  The model suite is outlined in Table \ref{tab:LdvsRs}.  Although moderate resolutions are sufficient for lower latitudes where the Rossby deformation length has values upwards of $1500$ km, high resolution is required to model polar dynamics where the value tends to decrease \citep{Read+2006a}.  In this suite, the lowest deformation length achieved was $750$ km.  The smallest possible perturbation that could be made on such a domain was $522$ km, ensuring that the perturbation radius was fully spatially resolved on the stencil.

\begin{deluxetable*}{ccccccccccccccc}
\tablecaption{Modeled parameter space for Case A runs shown in Figure \ref{f:LdvsRs}.  Here, the geopotential is maintained at $\langle gh \rangle = 200000$ m$^{2}$ s$^{-2}$, and all runs have \smax = 0.25 m$^2$ s$^{-3}$. \label{tab:LdvsRs}}
\tablewidth{0pt}
\tablehead{
\colhead{Model} &
\colhead{$L_d$} &
\colhead{$R_{\rm storm}$} &
\colhead{$\Omega$} &
\colhead{Bu$_l$} &
\colhead{Bu} 
&& && 
\colhead{Model} & 
\colhead{$L_d$} & 
\colhead{$R_{\rm storm}$} & 
\colhead{$\Omega$} &
\colhead{Bu$_l$} &
\colhead{Bu} \\
\colhead{\#} & \colhead{(km)} & \colhead{(km)} & \colhead{$\times\, \Omega_J$} & \colhead{-} & \colhead{$\times10^{-4}$} && && \colhead{\#} & \colhead{(km)} & \colhead{(km)} & \colhead{$\times\, \Omega_J$} & \colhead{-} & \colhead{$\times10^{-4}$}
}
% \decimalcolnumbers
\startdata
A1a & 750  & 522  & 1.69  & 2.06 & 0.96 && && A5b & 1750 & 750  & 0.73 & 5.43 &  5.24 \\   
A2a & 1000 & 522  & 1.27  & 3.67 & 1.71 && && A5c & 1750 & 1000 & 0.73 & 3.07 &  5.24 \\   
A2b & 1000 & 750  & 1.27  & 1.78 & 1.71 && && A5d & 1750 & 1250 & 0.73 & 1.96 &  5.24 \\   
A3a & 1250 & 522  & 1.02  & 5.72 & 2.67 && && A5e & 1750 & 1500 & 0.73 & 1.36 &  5.24 \\   
A3b & 1250 & 750  & 1.02  & 2.78 & 2.67 && && A6a & 2000 & 522  & 0.64 & 14.7 &  6.84 \\   
A3c & 1250 & 1000 & 1.02  & 1.56 & 2.67 && && A6b & 2000 & 750  & 0.64 & 7.11 &  6.84 \\   
A4a & 1500 & 522  & 0.85  & 8.26 & 3.85 && && A6c & 2000 & 1000 & 0.64 & 4.00 &  6.84 \\   
A4b & 1500 & 750  & 0.85  & 4.00 & 3.85 && && A6d & 2000 & 1250 & 0.64 & 2.56 &  6.84 \\   
A4c & 1500 & 1000 & 0.85  & 2.25 & 3.85 && && A6e & 2000 & 1500 & 0.64 & 1.78 &  6.84 \\   
A4d & 1500 & 1250 & 0.85  & 1.44 & 3.85 && && A6f & 2000 & 1750 & 0.64 & 1.31 &  6.84 \\   
A5a & 1750 & 522  & 0.73  & 11.3 & 5.24 && && \multicolumn{1}{l}{} & \multicolumn{1}{l}{} & \multicolumn{1}{l}{} & \multicolumn{1}{l}{} & \multicolumn{1}{l}{} & \multicolumn{1}{l}{} \\ \hline  
\enddata
\end{deluxetable*}

The suite shows the overall effect on nondimensional potential vorticity, $Q_e^*$ (equation \ref{e:Q_e}), of varying deformation length while keeping the injected pulse size fixed for each column (Figure \ref{f:LdvsRs}).  Deformation length $>1500$ km (and $1250$ km to a lesser extent) with the smallest perturbation exhibits the formation of dominant polar cyclones, as expected since the deformation length is directly related to the planetary Burger number.  Large values of $L_d$ produce Saturn-like behavior, where large scale polar cyclones dominate most of the flow, supporting the findings of \citetalias{Brueshaber+2019}.

In simulation A1a, (where $L_d=750$ km), turbulent self-organization of the emergent features results in a far more incoherent dynamical appearance as compared to the dominant polar cyclones at high $L_d$.  The deformation length modulates the proximity at which such storms can remain stable in each other's presence.  With perturbations set to the smallest of scales, the flow field can evolve the mass pulses and allow them to interact, coalesce and stabilize over time.  A small deformation length prevents them from aggregating, so most of the small storms that emerge remain relatively small and coherent.  Those that grow beyond the deformation length are affected by the Coriolis force.  The potential vorticity homogenizes over the domain and a cyclonic pattern similar to an $m=3$ wave emerges in simulation A1a.  Although this feature evolves in time due to the forcing model, the overall behavior remains the same.  We note that there is a dependence on the changing zonal wavenumber on deformation lengths.  This is seen very clearly in the movies where the wave number behavior emerges early on in the simulations \citep{hyder_ali_2022_6642986}.  For large planetary Bu, the low wave number begins to dominate much earlier in the evolution than low planetary Bu cases. In \S \ref{ssec:Char_L} below, we explain how the main zonal wavenumber emerging in our different simulations is controlled by the size of the largest balanced vortex that emerges from our forcing in each case.

Figure \ref{f:LdvsRs} shows that simulations with $L_d=1500$ km and above (A4a, A4b, A5a, A5b, etc.) exhibit more intense cyclonic features in the domain due to increased deformation length and increased perturbation size.  The maximum dimensional potential vorticity from the suite reaches $\sim 4\times10^{-3}$\,s\,m$^{-2}$, which is much higher than other models.  However, potential vorticity is expected to increase at the pole \citep{Showman_2007}, and these simulations represent an extreme parameter space (high perturbation radii).  The peak amplitude of the mass injection is the same throughout the suite, but varying \Rstorm~varies the FWHM of the Gaussian injection.  Larger values for \Rstorm~input significantly more energy into the system.  This drives stronger winds and causes a higher degree of forcing on the overall domain.  Although the winds are realistic, the emergent morphological features are different from what is observed on Jupiter's poles.  Although such values of \Rstorm$/L_d$ do not yield Jupiter-like dynamics, they are similar in appearance to cyclonic folded-filamentary regions observed on Jupiter across various latitudes, which are discussed in more detail in \S \ref{ssec:FFRcontent}.

It is worth noting that solely employing small \Rstorm$/L_d$ values is insufficient to produce a Jupiter-like pole, where the vortices are stable and do not merge.  Indeed, simulation A6a (where $L_d = 2000$ and Bu$_l\sim15$) produces a singular polar cyclone that dominates most of the polar region upwards of $80^{\circ}$.  This behavior breaks down with decreasing $L_d$ values.  Therefore, to approach a realistic Jovian pole, one must model increasingly small deformation lengths with varying injected pulse size.  This mechanism is not exhaustive as other parameters certainly play a role in the emergence of vortex configurations on Jupiter and their overall stability against the perpetual forcing from deeper atmospheric convection as shown by \cite{Cai2021DeepCP}.

\subsection{Folded-Filamentary Regions} \label{ssec:FFRcontent}

Jupiter's midlatitudes are dominated by anticyclones.  Most of the cyclonic features are expected to travel polewards via the Beta-Gyre Effect (or Beta Drift) \citep{Scott}.  However, cyclonicity can survive at low latitudes in the form of folded-filamentary regions (FFRs) and brown barges.  FFRs are elongated cyclonic structures that exist throughout the Jovian atmosphere and were first modelled numerically in \cite{Marcus2004}.  In that work, \cite{Marcus2004} notes that clouds of elongated forms are consistent with long-lived cyclonic behavior, and are dissimilar to clouds that form in anticyclonic regions.

There is observational evidence for such cyclonic filamentary structures at the high latitudes in JunoCam images returned by the Juno spacecraft \citep{Orton+2017}.  Our shallow water model successfully shows the emergence of FFRs at the high latitudes, given certain shearing conditions above \Rstorm\ $ \approx 1200$ km.  To our knowledge, this is the first such model to produce high latitude cyclonic FFRs and quantify a length scale beyond which such structures tend to emerge.  Formation of FFR-like \textit{beta-skirts}, which are regions of elevated potential vorticity surrounding a primary vortex \citep{Montgomery2006}, are observed in the models by \cite{ONeil+2015} and \citetalias{Brueshaber+2019} at lower Bu.  They note that such features tend to emerge in the beta-skirts surrounding the primary cyclones.  In our models, although elevated beta-skirts that exhibit filamentary behavior are seen around primary cyclones in the low $L_d$ cases, large scale FFR formation with significantly high mean horizontal shear is seen solely for the high deformation lengths with a perturbation scale beyond $1200$ km.  Furthermore, the high \Rstorm\ FFR-like features tend to be the main cyclonic features of those simulations rather than secondary features emerging in the vicinity of a cyclone.

Figure \ref{f:LdvsRs} shows the normalized form of non-dimensional potential vorticity.  It captures the skew-symmetric component of the velocity gradient tensor as it contains the local vorticity.  However, to understand the dynamical behavior of FFRs that emerge in the high $L_d$ models, we use the rate of strain tensor, which captures the symmetric component of the velocity gradient tensor and takes the form
\begin{equation}
S_{i,j} = \frac{1}{2}\left( \frac{\partial u_i}{\partial x_j} + \frac{\partial u_j}{\partial x_i} \right),
\end{equation}

\noindent where $i, j$ represent the Cartesian coordinates.  Physically, the rate of strain tensor represents the elongation of fluid elements and exhibits the total effect of local divergence and shear.  We compute the magnitude of the tensor using the Frobenius norm \citep{Golub}:
\begin{equation} \label{eq:frobenius}
\mid {\bf{S}} \mid = \sqrt{\sum_{i=1}^{m}\sum_{j=1}^{n} \mid S_{i, j} \mid ^2}.
\end{equation}

% \explain{In the above equation, the Sij term needed to be squared.  This was a typo - the calculation was done correctly.}

\noindent where $m=n=2$, as our models are limited to two dimensions. The norm of the rate of strain tensor encapsulates the combined effect of divergence that occurs where the FFRs fold over, and shear that occurs where fluid is deformed due to elongation.

\begin{figure*}[hbtp] \center
    \includegraphics[width=\textwidth]{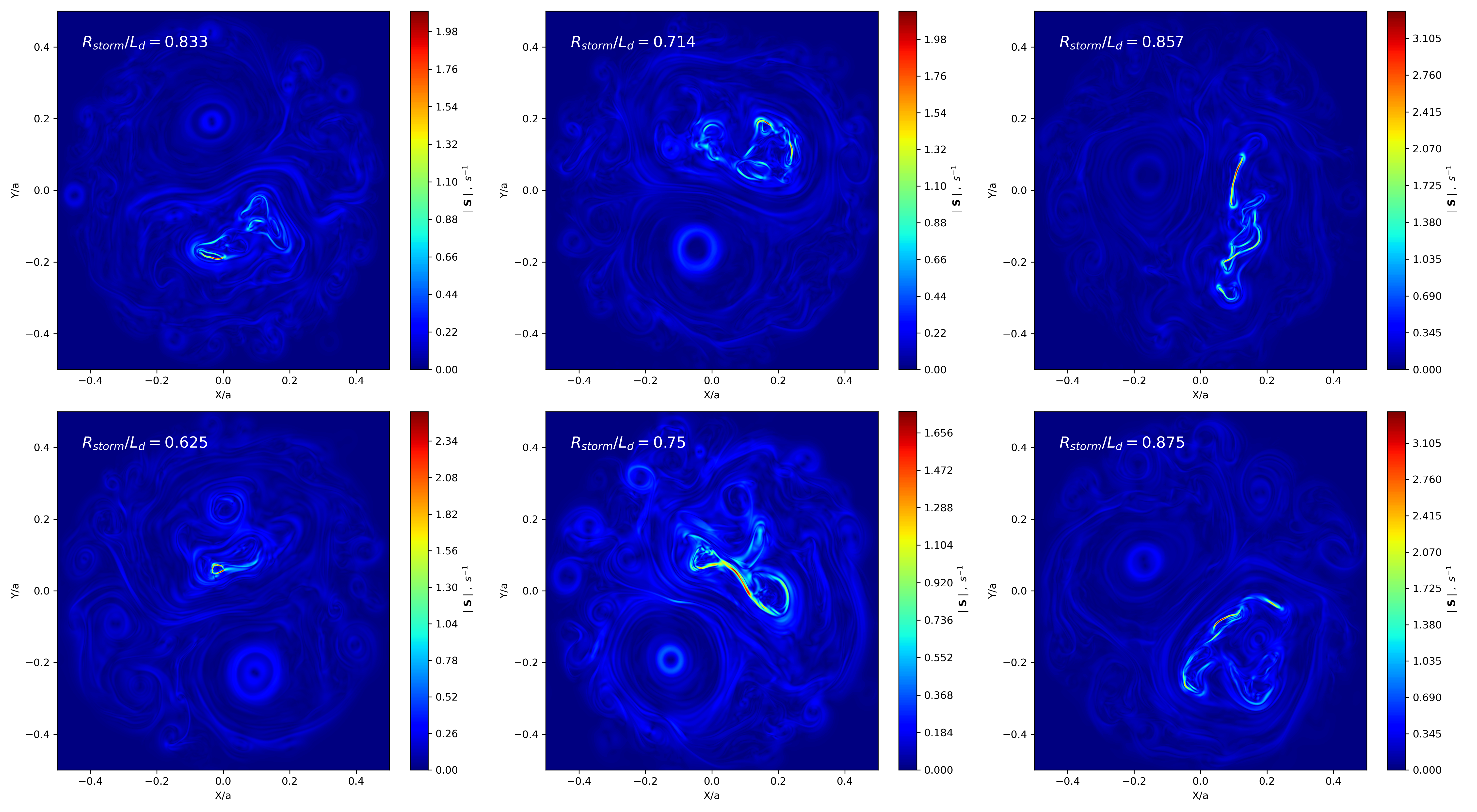}
    \caption{Rate of strain tensor magnitudes for models A4d, A5d, A5e, A6d, A6e, and A6f.  The strain rates are significantly higher at these regions relative to their environment.}
    \label{f:FFR_Smat}
\end{figure*}

\begin{figure*}[hbtp] \center
    \includegraphics[width=5in, height=4in]{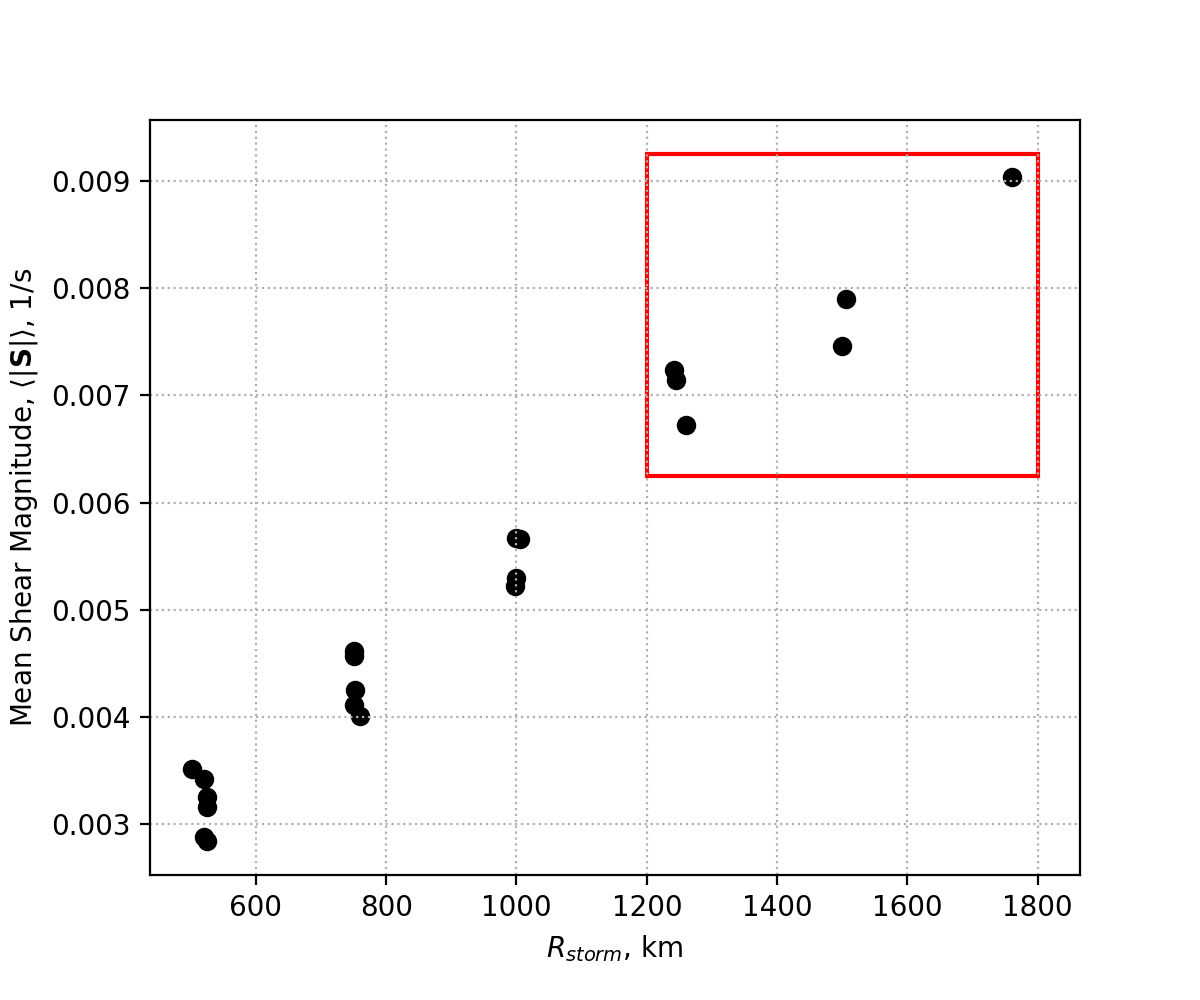}
    \caption{Mean shear magnitude at equilibrium for Case A simulations.  The points within the red rectangle pertain to the six simulations from Figure \ref{f:FFR_Smat} that exhibit FFR-type behavior.}
    \label{f:shearcorr}
\end{figure*}

\begin{deluxetable*}{ c  c  c  c  c  c  c }[]
\tablecaption{Dynamical quantities of interest for models that exhibit FFR-type behavior as shown in Figure \ref{f:FFR_Smat}.  The angle brackets denote a spatial average over the domain.} \label{tab3:FFR}
\centering
\tablewidth{0pt}
\tablehead{
\colhead{Model} & 
\colhead{$\textit{v}_{\rm{max}}$} & 
\colhead{$\langle v \rangle$} & 
\colhead{${(\nabla \cdot {\bf{v}})}_{\rm{max}}$} & 
\colhead{$\langle \mid\nabla \cdot {\bf{v}}\mid \rangle$} & 
\colhead{$\mid{\bf{S}}\mid_{\rm{max}}$} & 
\colhead{$\langle \mid{\bf{S}}\mid \rangle$} \\
\colhead{\#} & 
\colhead{ms$^{-1}$} & 
\colhead{ms$^{-1}$} & 
\colhead{$\times10^{-2}$\ (s$^{-1}$)} & 
\colhead{$\times10^{-4}$\ (s$^{-1}$)} & 
\colhead{(s$^{-1}$)} & 
\colhead{$\times10^{-3}$\ (s$^{-1}$)}
}
\startdata
A4d & 121 & 23 & 6.953 & 1.680 & 0.432 & 7.147  \\
A5d & 137 & 27 & 3.329 & 1.890 & 0.289 & 7.237  \\
A5e & 156 & 29 & 6.640 & 2.996 & 0.467 & 7.899  \\
A6d & 143 & 28 & 9.691 & 1.877 & 0.360 & 6.720  \\
A6e & 152 & 32 & 2.038 & 2.278 & 0.220 & 7.464  \\
A6f & 189 & 35 & 4.344 & 3.004 & 0.352 & 9.039 
\enddata
\end{deluxetable*}

Figure \ref{f:FFR_Smat} shows a number of interesting features.  The plot shows the rate of strain tensor magnitudes using equation \ref{eq:frobenius} of a subset of models from Figure \ref{f:LdvsRs}, which reflect the behavior of increased \Rstorm.  The strain-rates are able to capture the high shear behavior in folded regions well.  Such FFR-like behavior is morphologically distinct from FFRs in beta-skirts around cyclones, as the FFRs in our simulations represent the primary mode of cyclonicity in the flow.  The corresponding divergence, shear, and maximum wind speed values are presented in Table \ref{tab3:FFR}.

Although cyclonic filamentary regions have been observed at the high latitudes by \textit{Juno}, their underlying wind speeds remain unconstrained.  Thus, we limit the emergent mean wind speed of models with such elongated structures to the observed zonal speeds of the ambient winds and storms, ensuring that we remain under the overforcing regime.  The strength of the forcing, and more precisely, the injected geopotential relative to the total geopotential, may have a significant effect on the emergence of such morphology.  Our model shows the formation of FFRs, but is strongly dependent on the mass injection size (perturbation radii).  Figure \ref{f:shearcorr} supports the idea that emergent FFRs are regions of dominant shear.  The presence of such features at Jupiter's polar regions as seen in \textit{Juno} observations suggests that FFRs form either due to intense vortex interactions with large moist convective events enabled due to the planetary vorticity gradient, or they emerge due to other mechanisms not captured by the shallow water approximation.

\subsection{Forcing Strength} \label{ssec:forcingstrength}

\begin{figure*}[hbtp] \center
    \includegraphics[width=\textwidth]{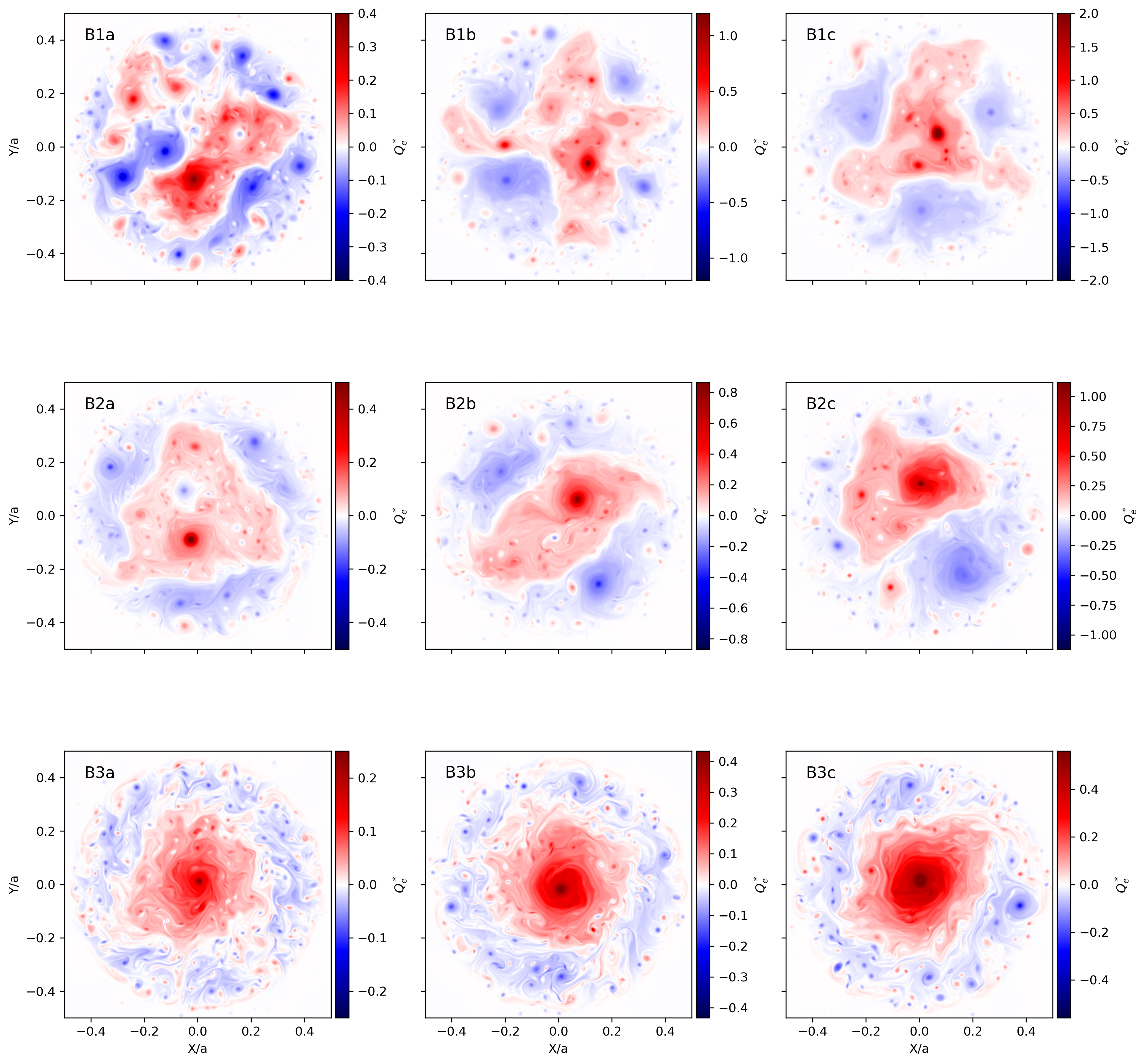}
    \caption{$Q_e^*$-dependence on forcing strength and deformation length.  \textit{Top to bottom:} Increasing deformation length, $L_d = \{750, 1250, 2000\}$ km. \textit{Left to right:} Increasing geopotential ratio, $\Phi_r = \{7\%, 14\%, 20\%\}$.  Movies for these Case B simulations are available on Zenodo: \dataset[doi:10.5281/zenodo.6642986]{https://doi.org/10.5281/zenodo.6642986} \citep{hyder_ali_2022_6642986}. Similar to Case A, the movies show a stronger dependence on deformation length compared to the forcing strength.  However, for smaller deformation lengths the effect of forcing strength is much larger.}
    \label{f:deltatest}
\end{figure*}

\citetalias{Brueshaber+2019} and \cite{Brueshaber+2021} found that storm energetics did not play an important role in the equilibrated dynamical state of the system.  In their models, they used a fixed geopotential (our Case A).  Here, we use Jupiter's observed rotation rate (Case B), and vary the forcing strengths that may be relevant to the Jovian pole in particular.  In order to vary the injected geopotential relative to the total geopotential, we use $\Phi_r$, as defined in equation \ref{eq:Delta}.  Deformation lengths of $750$, $1250$, and $2000$ km were used for this model suite.  For each choice of $L_d$ {$(Bu)$}, we varied the injection strength such that $\Phi_r \approx 7\%, 14\%$, or $20\%$.  The simulations are presented in Figure \ref{f:deltatest} and the parameters are provided in Table \ref{tab:forcingstrength}.

\begin{deluxetable*}{ccccccccccccccc}
\tablecaption{Modeled parameter space for Case B runs as shown in Figure \ref{f:deltatest}.  Here, the storm strength is varied while Bu$_{l}$ is kept constant for the choice of Bu. \label{tab:forcingstrength}}
% \tablewidth{0pt}
\tabletypesize{\footnotesize}
% \resizebox{\textwidth}{!}
\tablehead{
\multicolumn{3}{c}{Bu$=9.6\times10^{-5}$}
&& &&
\multicolumn{3}{c}{Bu$=2.67\times10^{-4}$}
&& &&
\multicolumn{3}{c}{Bu$=6.84\times10^{-4}$}
\\
\hline
\multicolumn{3}{c}{Bu$_l=2.06$}
&& &&
\multicolumn{3}{c}{Bu$_l=5.72$}
&& &&
\multicolumn{3}{c}{Bu$_l=14.7$}
\\
\hline
\colhead{Model} &
\colhead{\smax} &
\colhead{$\Phi_r$}
&& &&
\colhead{Model} &
\colhead{\smax} &
\colhead{$\Phi_r$}
&& &&
\colhead{Model} & 
\colhead{\smax} & 
\colhead{$\Phi_r$}
\\
\colhead{\#} & \colhead{(m$^2$ s$^{-3}$)} & \colhead{\%} && && 
\colhead{\#} & \colhead{(m$^2$ s$^{-3}$)} & \colhead{\%}&& && 
\colhead{\#} & \colhead{(m$^2$ s$^{-3}$)} & \colhead{\%}
}
% \decimalcolnumbers
\startdata
B1a & 0.028 & 7  && && B2a & 0.079 & 7  && && B3a & 0.202 & 7  \\
B1b & 0.057 & 14 && && B2b & 0.158 & 14 && && B3b & 0.404 & 14 \\
B1c & 0.085 & 20 && && B2c & 0.237 & 20 && && B3c & 0.606 & 20 \\
\hline  
\enddata
\end{deluxetable*}

We note that the main differences in the observed variations in dynamical behavior occur due to the deformation length.  However, the storm strengths have an important impact on the low deformation length models, particularly with regard to how potential vorticity homogenizes over the domain.  As expected, simulation B1a produces low values for $Q_e^*$ and results in low overall wind speeds.  Higher wind speeds are produced for higher $\Phi_r$ cases.  Although all simulations show the formation of a polar cyclone to some degree, the cyclonic beta-skirts of the low $L_d$ simulations, in general, homogenize distinctly from the high $L_d$ cases.  This holds true for all values of $\Phi_r$.  The cyclonic potential vorticity aggregates at the poles as expected, but it tends to form an almost triangular pattern, similar to the $m=3$ pattern seen in low $L_d$ models in Case A.  Although, the final snapshot of simulation B1a shows an almost bimodal structure in $Q_e^*$, it also exhibits the $m=3$ wave behavior throughout its evolution.  If an anticyclonic region is surrounded by cyclonic fluid, as is seen in the final snapshot of simulation B2a, the anticyclonic patch is ejected over time due to the beta-gyre effect, leaving a net accumulation of cyclonicity at the pole.

\begin{figure*}[hbtp] \center
    \includegraphics[width=\textwidth]{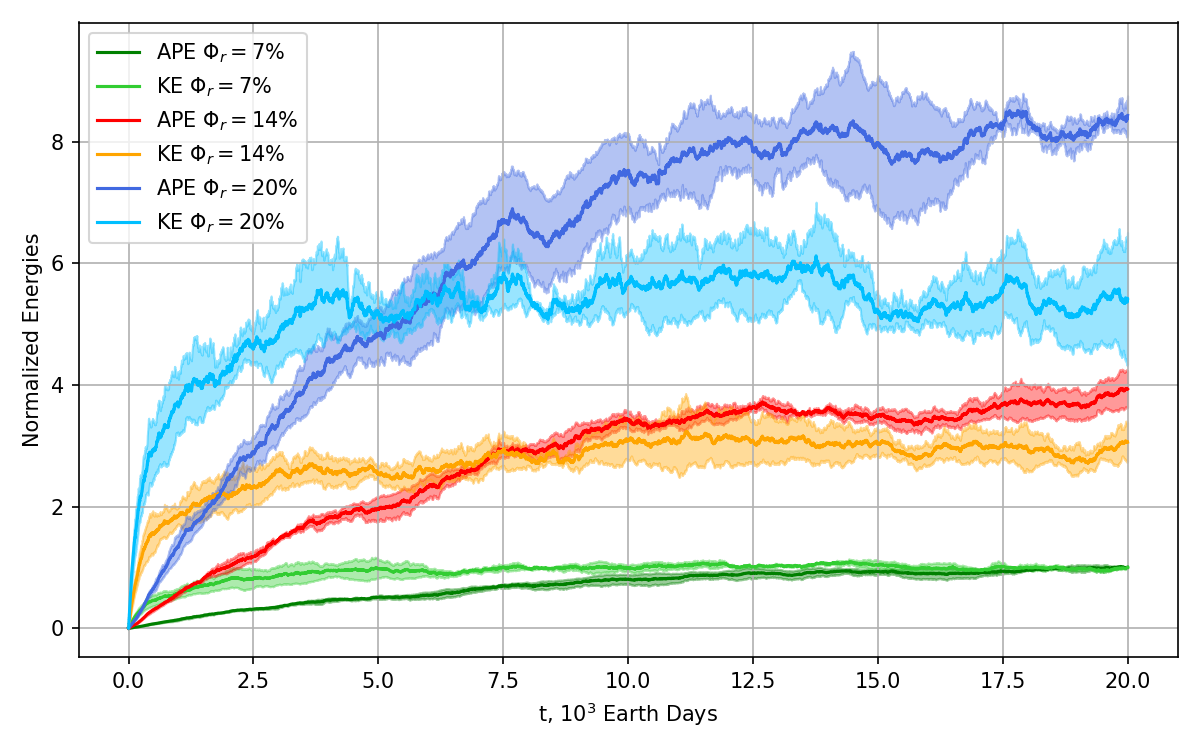}
    \caption{Available potential energy (APE) and kinetic energy (KE) for simulations in Case B, where $L_d$ has been averaged over.  The energies have been normalized by the final values of the $\Phi_r=7\%$ simulations.  The shaded regions show the $1\sigma$ variation across the various deformation lengths.  Changing $\Phi_r$ has a significant effect on when the APE becomes dominant over KE.}
    \label{f:deltaenergies}
\end{figure*}

Although the dynamics evolve in time, the energy stabilizes relatively early in the simulations.  Shallow water available potential energy (APE) and kinetic energy (KE) are defined as
\begin{eqnarray}
\label{e:APE}
APE &=& \frac{1}{2} \int (gh)^2 - {\langle gh \rangle}^2\ dA,\\
\label{e:KE}
KE  &=& \frac{1}{2} \int gh(u^2 + v^2)\ dA,
\end{eqnarray}

\noindent where $dA$ is an areal element of the domain, and the integrals are taken over the entire box.  The APE tends to be an order of magnitude higher than the kinetic energy even as the simulations equilibrate.  Figure \ref{f:deltaenergies} shows the energies relevant to Case B.  The energies have been normalized by the final values of the $\Phi_r=7\%$ simulations, and averaged over $L_d$.  The shaded regions show the $1\sigma$ variation that is due to changing deformation lengths.  In terms of the energy, we note that the effect of varying forcing strength is much larger than varying deformation length.  $\Phi_r$ directly affects the KE to APE conversion point.  Simulations with stronger forcing tend to convert their KE into APE earlier in their evolution, whereas simulations with weak forcing exhibit a less efficient conversion between KE and APE.  This can be seen in the Case B movies where large planetary Bu simulations manifest the low wave number behavior early in their evolution compared to the low planetary Bu cases \citep{hyder_ali_2022_6642986}.  The models stabilize by $10^4$ Earth days in KE while APE tends to equilibrate more slowly, in agreement with \citet{Showman_2007}.  APE remains about an order of magnitude higher than KE.  Our APE remains above KE by a larger factor than the low latitude shallow water study performed by \citet{Showman_2007}, but agrees with polar modeling results from \citetalias{Brueshaber+2019}.

\label{ssec:Char_L}
\subsection{Evolution of the Characteristic Length}

\begin{figure*}[hbtp] \center
    \includegraphics[width=\textwidth]{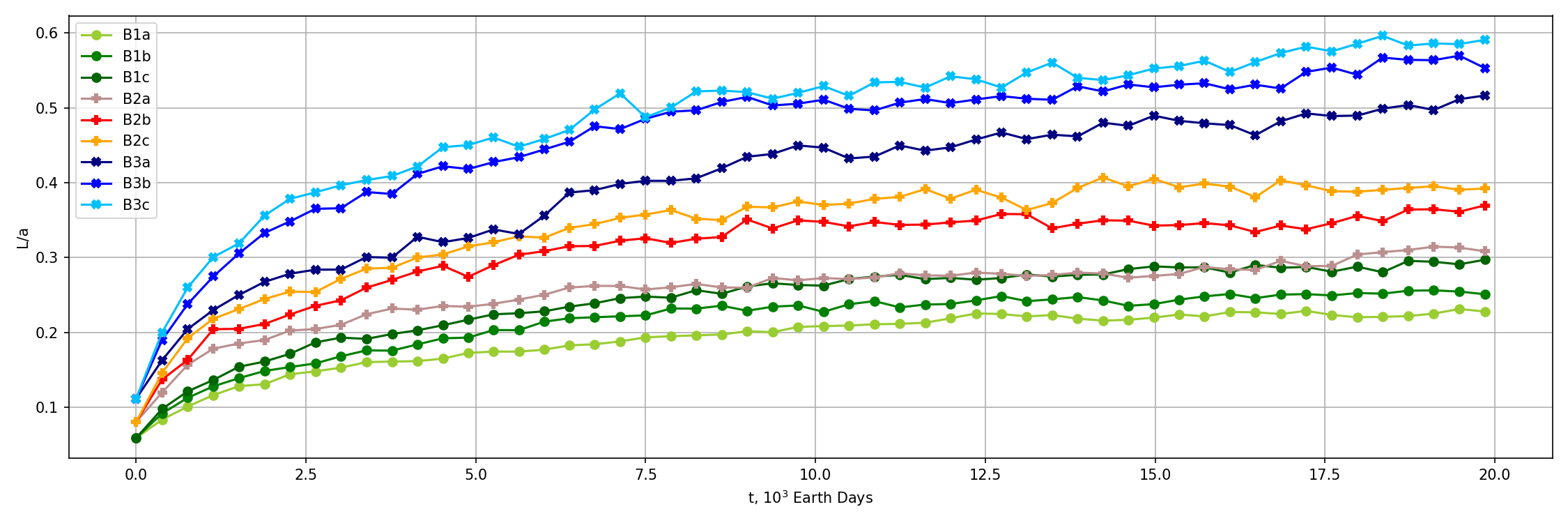}
    \caption{Characteristic length of the dominant balanced vortex, $L/a$, plotted as a function of time for all Case B simulations.  The profiles plateau as the simulations stabilize their APE to KE ratios.}
    \label{f:charL}
\end{figure*}

The ratio of APE to KE is directly related to the length scale of emergent vortices via
\begin{equation} \label{eq:L/Ld}
\frac{APE}{KE} = C \left(\frac{L}{L_d}\right)^2,
\end{equation}

\noindent where $L$ is the size of the largest balanced vortex/eddy, and $C$ is a constant of proportionality \citep{Pedlosky1979, MarshallJohn2008Aoac}.  We determine $C$ by using simulation B3a.  The characteristic size of the largest balanced vortex emerging in this simulation is $\sim 0.55a$ (where $a$ is the domain size).  Using the final $APE/KE$ value for simulation B3a and $L\sim0.55a$, $C$ is found to be about $0.025$.  We show $L/a$ as a function of time for the simulations in Case B in Figure \ref{f:charL}.  Given $L$, the maximum number of eddies, $N$, in a domain with size $a$ would be proportional to $a^2/L^2$.
 
The wavenumber behavior observed in our simulations comes as result of eddies balancing out to occupy a circular domain.  It is dependent on how the $N$ vortices self-organize to minimize the total energy.  The $m=1$ behavior of simulations B3a, B3b, and B3c corresponds to how the dominant polar vortex centralizes in the domain, given $L/a\sim0.55$.  This leaves no room for any other major storm to share the domain. For simulations B1c and B2a, an $m=3$ pattern emerges that makes these two simulations look almost identical in the last simulated day. As seen in Figure \ref{f:charL}, these two simulations reach approximately the same value of $L/a\approx 0.3$ by the end of the simulation. This value is similar to the one that would be obtained from circle-packing arguments (L/a=0.333) \citep[see e.g.,][]{Pirl1969DerMV} for a total number of vortices equal to 7, which is the number of large balanced vortices seen filling the domain in Figure \ref{f:deltatest} (one in the center of the domain and six of alternating signs surrounding it). The same argument can be used to explain the other patterns that emerge from our simulations. For example, for simulations B1a and B1b the number of balanced vortices that emerge by the end of the simulations is $N\sim\{13, 9\}$, which from circle packing arguments would correspond to values of L/a of 0.236 and 0.277 respectively \citep{Pirl1969DerMV, Fodor2003}, while the values we obtain from our analysis are 0.225 and 0.25. While our L/a values are in good agreement with those derived from circle packing arguments, they are always slightly smaller, since the largest balanced vortices that emerge from our simulations cannot be tightly packed.

These same arguments can also be used to explain the patterns that emerge in the simulations in case A. The energy analysis for simulations A4a, A5a, and A6a reveals that the ratio of L/a$\sim0.5$, which would explain why most panels in Figure \ref{f:LdvsRs} look like there are only two large balanced vortices sharing the domain. The main difference with case B is that in this case none of these eddies occupy the center of the domain, therefore allowing for an additional dominant vortex, with a characteristic length scale of $\sim0.5$, to remain stable.

%For simulations B1a, B1b, and B1c, an $m=3$ pattern emerges.  These runs have $L/a=\{0.225, 0.250, 0.293\}$, respectively (Figure \ref{f:charL}).  From Figure \ref{f:deltatest}, we can see that the approximate number of balanced eddies in these simulations is $N\sim\{13, 9, 7\}$ resulting in $L/a=\{0.236, 0.277, 0.333\}$, respectively, from circle-packing arguments \citep[see e.g.,][]{Pirl1969DerMV, Fodor2000, Fodor2003}.  These are in good agreement with the measured values.  The predicted values are higher than those obtained using $N$ as the vortices that emerge in our simulations are not perfectly packed circles, but regions of increased vorticity with fluid bounds.  Thus, the expected number of balanced eddies will always be higher than the number of eddies that emerge.

% Therefore, the number of balanced eddies can be written as
% \begin{equation} \label{eq:numberofspots}
% N = C \left(\frac{a}{L_d}\right)^2 \frac{KE}{APE}.
% \end{equation}

\subsection{Dynamical Instability} \label{ssec:instability}

Dynamical instabilities are a consequence of a system's inability to withstand perturbations to the mean flow.  As small perturbations grow in an unstable system, they tend to have long term effects on the overall behavior of the flow \citep{Chandrasekhar1961}.  In forced-dissipative simulations, such perturbations continuously perturb the system, resulting in variations to the underlying stability.  As our quasi-2D model uses the shallow water approximation, there are no solenoidal contributions to the system that may excite baroclinic modes, thereby limiting the domain to barotropic instabilities exclusively.  We examine the long term behavior of our simulations to see if they are Arnol'd- stable.

\begin{figure*}[hbtp] \center
    \includegraphics[width=\textwidth]{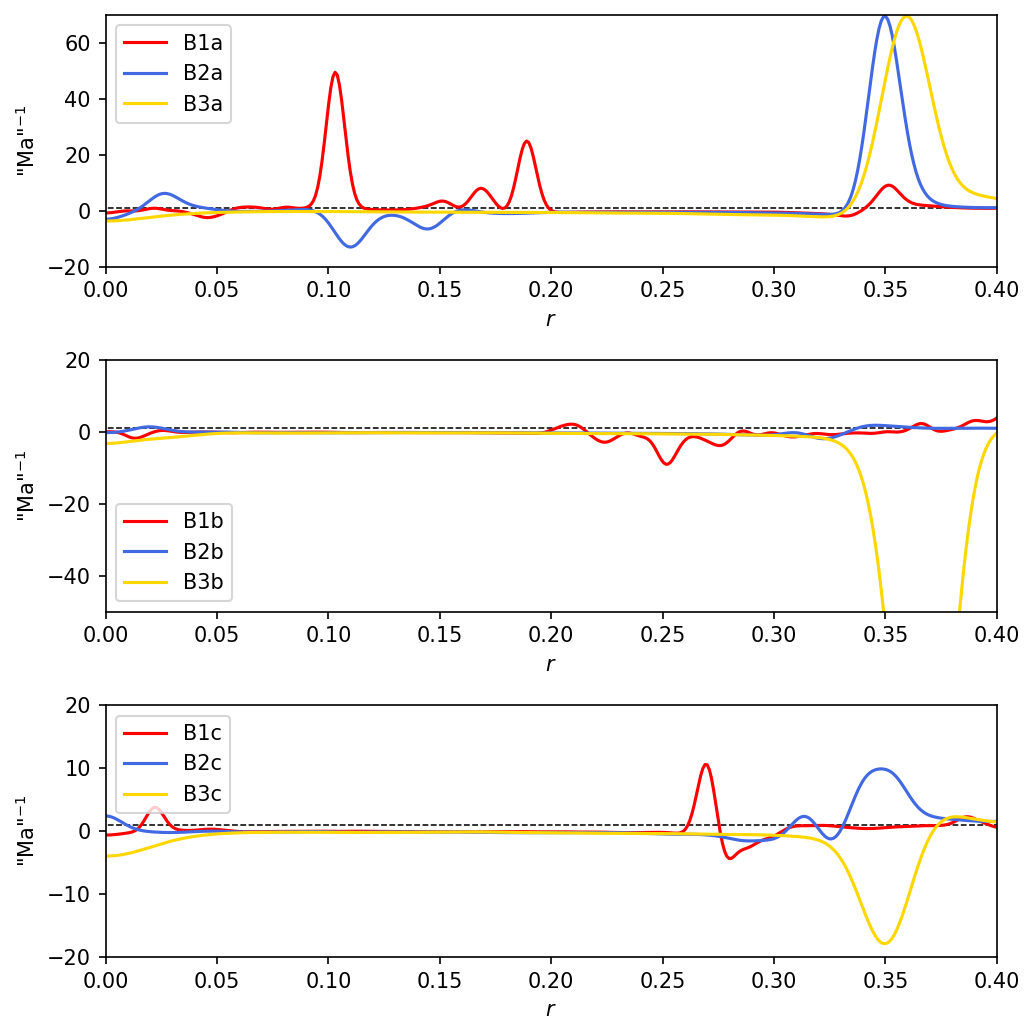}
    \caption{``Ma"$^{-1}$ as a function of distance from the pole for all simulations used in Case B.  The dashed black line demarcates ``Ma"$^{-1}=1$.}
    \label{f:arnold}
\end{figure*}

The Arnol'd stability theorems utilize the pseudoenergy, $H$, which is a combination of the kinetic energy and enstrophy (squared vorticity) and is defined as
\begin{equation}
\begin{split}
H & \equiv \frac{1}{2} \int \left[(\nabla \psi')^2 + \frac{d\Psi}{dQ}(q')^2 \right]\ dA,
\end{split}
\label{e:pseudodenergy}
\end{equation}

\noindent where $\Psi$ and $Q$ represent the geophysical steady-state stream function \citep{McintyreShepherd1987} and absolute vorticity, respectively.  Correspondingly, $\psi'$ and $q'$ are their perturbed values \citep{VallisBook}.  Stability of the flow requires that the pseudoenergy is positive-definite or negative-definite, thus the emergence of instabilities rests on the behavior of $d\Psi/dQ$.  Arnol'd's first stability condition requires that $d\Psi/dQ > 0$, while the second requires that it be sufficiently negative such that $H$ remains negative-definite across the domain.  \citet{Dowling1993} showed that if $d\Psi/dQ < -L_d^2$, Arnol'd's second criterion is sufficiently satisfied.  In our case, we may write
\begin{equation}
\begin{split}
\frac{d\Psi}{dQ} & = \frac{d\Psi/dr}{dQ/dr} = -v_{\theta} \left(\frac{d\zeta}{dr} + \frac{df}{dr}\right)^{-1},
\end{split}
\label{e:arnoldmodelderiv}
\end{equation}

\noindent where $v_{\theta}$ is the zonal wind.  \citet{Dowling2020} showed that the two Arnol'd-stable branches can be concatenated in to a singular condition using the inverse of the ``vortical" Mach number, ``Ma"$^{-1}$, where ``Ma"$^{-1} < 1$ implies an Arnol'd-stable flow.  The vortical Mach number is defined as.
\begin{equation}
\begin{split}
``\rm{Ma}"^{-1} = -L_d^2\ \frac{dQ}{d\Psi}.
\end{split}
\label{e:InvMach}
\end{equation}

\citet{Andrews1984} found that Arnol'd-type stability can only occur for zonally symmetric cases.  However, \citet{Carnevale+1990} and \citet{MuWu2001} provided further extensions to the theorem, although the restrictions are non-trivial \citep{Read2020}.  Here, we only include this analysis as a means to study the stable or unstable behavior of the longterm trends in our simulations.  Limiting the applicability to satisfy the results of \citet{Andrews1984} means that the analysis can only be applied to high deformation length cases (simulations B3a, B3b, and B3c).  However, for completion, we provide the results for all Case B simulations to highlight the distinct profiles of ``Ma"$^{-1}$.

We use the steady-state velocity and vorticity fields over the last 5000 days.  This results in azimuthally symmetric fields for most of the Case B simulations. We then take an azimuthal average to produce a radial profile for ``Ma"$^{-1}$.  Since we are interested in the stability of the emergent structures that are affected by the choice of deformation length and not the perturbations at the grid scale, we convolve the profile with a Gaussian kernel with a full width half maximum (FWHM) set to the corresponding $L_d$ for each simulation.

Figure \ref{f:arnold} shows the profile of ``Ma"$^{-1}$ as a function of distance from the pole, $r$.  By day 20,000, all dynamics have equilibrated (see Figure \ref{f:deltaenergies}).  Simulations B3a, B3b, and B3c show that the polar cyclones are Arnol'd-stable out to at least $r\sim0.3$, beyond which the ``Ma"$^{-1}$ profiles cross the ``Ma"$^{-1}=1$ transition.  The profiles show stability where the polar cyclones dominate.  There is a noticeable dependence on the forcing strength, which is clear in simulations B3a, B3b, and B3c.  As the forcing strength is increased, the crossing point is pushed equatorward from $r\sim0.3$ (B3a) to $r\sim0.4$ (B3c).  This is because as the forcing strength increases, the polar cyclone gets stronger (Figure \ref{f:deltatest}) resulting in a larger domain of stability.  The interpretations of other simulations in Case B are difficult with regards to Arnol'd-stability.  However, the distinction between the high and low $L_d$ cases is clear.  Lower $L_d$ simulations show less homogeneous behavior in their PV fields, which is reflected in their ``Ma"$^{-1}$ profiles.

\section{Discussion} \label{sec:furtherdisc}

Our model provides insight into the emergence of FFRs, application of moist convective theory, and an exploration of the low deformation length regimes in gas giant atmospheres. Here we discuss the primary results from our simulations in context of current modeling efforts.

\subsection{Emergence of FFRs}

Our simulations show the emergence of FFRs at high \Rstorm\ values, indicating that a larger turbulent forcing scale, which may inject more localized available potential energy, is responsible for higher values of horizontal shear (see Figure \ref{f:shearcorr}).  Even though not all of the injected energy is able to convert to kinetic energy, a large turbulent forcing scale injects more geopotential in a region at the same forcing strength, \smax.  \citet{Marcus2004} showed that twisted filamentary structures are consistent with long-lived cyclonic behavior.  We observe that the high horizontal shear in our simulations is exclusively localized to the cyclonic regions of the domain, in agreement with their result.

\begin{figure*}[hbtp] \center
    \includegraphics[width=\textwidth, height=3.5in]{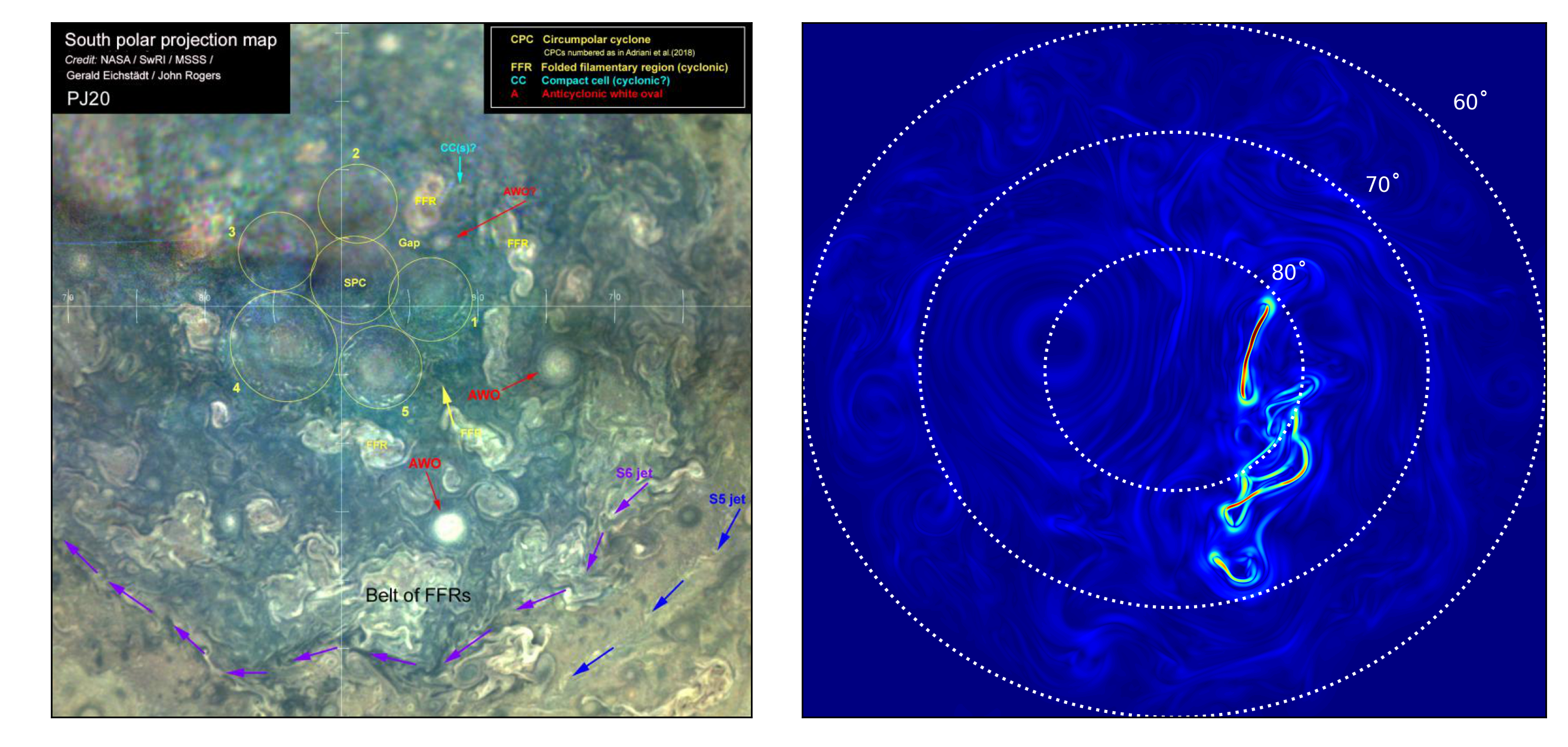}
    \caption{\textit{Left}: JunoCam RGB processed image of the South Polar Region at PJ20 \citep[][image credit: NASA/SwRI/MISSS/Gerald Eichst{\"a}dt/John Rogers]{ROGERS2022114742}.  The FFRs (demarcated in yellow) are offset from the pole and constitute a belt of large scale FFRs beyond about $85^{\circ}$S.  \textit{Right}: Simulation A5e from Figure \ref{f:FFR_Smat}.  The FFR spans a large area in the domain and is also offset from the pole, similar to what is seen by JunoCam.}
    \label{f:junocam_FFR_comparison_A5e}
\end{figure*}

The FFR behavior tends to occur consistently above a specific forcing scale in our simulations.  Above \Rstorm$=1200$ km, horizontal shear begins to dominate over the divergence by about an order of magnitude (see Table \ref{tab3:FFR}).  Thus, the bulk of the available potential energy is converted to shear as the fluid element elongates and folds over.  This ultimately results in extremely high localized velocities, e.g. simulation A6f has $v_{\rm max} = 189$ m s$^{-1}$ and is well beyond any observed wind speed at the polar regions, however, the domain-averaged velocity magnitudes remain below $\sim 35$ m s$^{-1}$, in agreement with observations \citep{Orton+2017}.  Furthermore, some high latitude FFRs have been identified as sources of lightning, suggesting a moist convective origin \citep{Borucki1992, Fletcher2017}.  The formation of FFRs above a certain forcing scale suggests that large scale moist convective events might be the dynamical source of amorphous, filamentary cyclonicity at the poles as they deposit more APE to the upper atmosphere that can be converted to horizontal shear.  In Figure \ref{f:junocam_FFR_comparison_A5e}, we show a comparison of a JunoCam image of the South polar region during Perijove 20 (PJ20) with one of the simulations that contain a large scale FFR-type feature.  Qualitatively, the modeled FFR is larger than any of the observed FFRs in the JunoCam image.  However, the simulation exhibits a similar structure in its morphology, and position relative to the pole.  The belt of FFRs $(70^{\circ}-85^{\circ}$ S)\ hosts an assembly of FFRs that are similar in morphology to the large scale FFR behavior we see in our simulations.

\subsection{Barotropic Modes and Moist Convection}

Our model shows that exclusively employing barotropic modes in the simulations is sufficient to produce the dynamical variety of the polar atmosphere, consistent with previous works \citep{ONeil+2015, ONeill+2016, Brueshaber+2019, Brueshaber+2021}.  The consistency of these results provides strong support to the idea that moist convective behavior of the deep atmosphere functions as an important source of observed barotropic vortices of the higher latitudes.  Our findings lend support to the hypothesis put forth by \citet{ONeil+2015} that the planetary Burger number functions as the primary non-dimensional parameter that modulates the long term evolutionary behavior of polar cyclones.  It lends further credence to the findings of \citetalias{Brueshaber+2019}, {\it i.e.} that exploration of the parameter space by varying planetary Bu, injection strengths, and cyclone-to-anticyclone ratio provides dynamical variation that may be relevant to the atmospheres of gas and ice giants.  Indeed, this distinction is further explored in \cite{Brueshaber+2021}, which focused exclusively on the overall wind speeds and evolutionary structure of polar cyclones that may elucidate the primary contrast between Saturn-like (``S") and Ice Giant-like (``I") regimes in giant planets.

In Figure \ref{f:LdvsRs}, the low deformation length case (simulation A1a) exhibits behavior similar to the Jupiter-like (``J") systems and begins to exhibit more transitional (``T") morphology as $L_d$ is increased.  The intermediate $L_d$ simulations in Figure \ref{f:LdvsRs} show more of a homogenized distribution of cyclonic PV, however, no dominant polar cyclone emerges until an $L_d$ of $1500$ km is reached, where the rotational frequency becomes lower than $\Omega_J$.  However, this is also dependent upon the turbulent forcing scale of the model.  For simulations with intermediate $L_d$, increasing the injection scale results in the formation of a much more dominant polar cyclone, which is reminiscent of high $L_d$ cases.  Thus, there is a non-linear relation between the interplay of \Rstorm\ and $L_d$, which directly supports the findings of \citet{ONeil+2015}.  Both parameters modulate the flow via conversion of the injected APE into localized winds, and therefore KE, as well as imparting strong gravity waves.  The choice of the local Burger number, Bu$_l$, therefore, modulates how much of the injected potential energy undergoes geostrophic adjustment and how the resulting KE interacts with the ambient flow.  The model, however, is limited to barotropic modes as perturbations are made solely made to single layer geopotential field.  This sufficiently captures the effect of mass and energy changes due to vertical mass transport via moist convective events \citep{SanchezLavega2011, GarciaMelendo2013, GARCIAMELENDO2017}, but is not sufficient to reveal deep atmospheric behavior of the moist convective plumes themselves.  A baroclinic anomaly, as used by \citet{ONeil+2015} to simulate moist convective events, would result in vertical shearing of the hetons, leaving behind a strongly barotropic vertical structure \citep{Skinner2021}.  However, for a comprehensive approach to moist convection of stratified baroclinic anomalies, 3D models are necessary to constrain the deep atmospheric behavior and its effect on the upper troposphere \cite[see][for a full 3D treatment of deep atmospheric flows]{Garcia+2020}.

\subsection{Completion of Parameter Space - Exploring low planetary Burger number}

This work introduces an accurate and high resolution model that is necessary to simulate the high latitude dynamics on gas giants, particularly Jupiter.  Providing an extension of the parameter space setup in \citetalias{Brueshaber+2019} and \citet{Brueshaber+2021}, our model is able to resolve the lower end of the deformation length parameter that functions as the primary variant that modulates the planetary Burger number.  Although \citetalias{Brueshaber+2019} explores similar Bu regimes, their model is restricted to extremely large planetary scales.  The benefit of addressing the case for Jupiter explicitly supports the applicability of the moist convective theory to the Jovian high latitudes, and due to computational capabilities afforded by the Pencil framework, allows for a much richer picture of the shallow dynamics that may be driven by moist convective events occurring in the deeper atmosphere.

We limited our work to a resolution of $1024^2$ in order to produce the main aspects of the parameter space that is explored by \citetalias{Brueshaber+2019}.  Furthermore, \citet{Brueshaber+2021} expand upon the details of the ``S" and ``I" dynamical regimes to furnish the details of polar cyclone behavior, which correspond to Bu $\sim1.6\times10^{-3}$ and Bu $\sim10^{-2}$, respectively.  Here, we provided a necessary investigation of the Jovian ``J" dynamical regime $({\rm Bu} \sim 10^{-5})$.  Our findings support the idea that barotropic modes are sufficient to capture the dynamical variety observed on Jupiter's polar regions, even though the simulations are not able to produce the vortex configurations \citep{Adriani+2018}.  The vortex configurations require a balance between the meridional vorticity gradient generated by the polar cyclone and the planetary vortcity gradient.  This balance results in a region of stability that is able to host multiple circumpolar cyclones, as long as they remain at a distance of a few $L_d$ from each other in order to minimize disruption \citep{Gavriel2021}.  Our simulations successfully show the emergence polar cyclones, even in the low $L_d$ cases, but are unable to produce similar sized circumpolar structures, suggesting that the balance between the polar cyclone vorticity gradient and the planetary vorticity gradient is easily disrupted by continuous forcing.  Using a freely decaying model, \citet{Li+2020} showed that once the cyclones are stabilized against the planetary vorticity gradient, initialized vortex configurations evolve in equilibrium.  In a forced turbulence model such as ours, meridional vorticity gradient balance is continuously disrupted and no configuration is able to form.

\subsection{Shallow Water Applications of the {\sc{Pencil Code}}}

This work applies the computational framework of Pencil to the atmosphere of gas giant planets for the first time.  The code is open source and allows for users to model a wide variety of astrophysically relevant dynamics and thermochemistry in with modern computational techniques.  This allows for increased time savings due to significantly faster model integration.  Specifically, the code allows for full $20,000$ days of evolution for the lowest perturbation size and lowest deformation length simulation, A1a (most computationally expensive simulation due to relevant size scales), within a wall-clock time of $45$ hours.  Our model also requires the turbulent forcing scale to be resolved fully by the stencil, a condition that required far more wall-clock time in previous works (e.g. \cite{Brueshaber+2021}).

Our model produces results similar to those of previous works and further adds to the increasing body of knowledge that motivates 3D modeling of the atmosphere in the future \citep{Garcia+2020}.  We are able to show distinct and realistic morphologies that emerge due to a consequence of our parameterizations using the Pencil Code.  Although we maintain the resolution at $1024^2$, we test the code for convergence up to $2048^2$.  Thus, the parameter space can be expanded further with the inclusion of higher resolution simulations and deformation lengths as low as $500$ km.

\section{Conclusions}\label{sec:conclusion}

In this work, we showcase a novel application of a well-tested hydrodynamical code to the atmosphere of gas giant planets capable of addressing wide parameter space searches by utilizing the framework provided by {\sc{PencilCode}}.  Applying this atmospheric module to the case of Jupiter's polar regions using high resolution, we find high latitude behavior reminiscent of FFRs.  The simulations show that these FFRs are regions of intense horizontal shear and are generated by storms above a length scale of $\sim 1200$ km.  We further explore an immense parameter space defined by \citet{ONeil+2015}, \citetalias{Brueshaber+2019}, and \citet{Brueshaber+2021}, and produce results that support the idea that moist convection functions as a dominant source of the dynamical variety that is seen at the Jovian poles.

We find that the deformation length, $L_d$, functions as a primary indicator of long term polar behavior, and specifically that increasing the turbulent forcing scale for large $L_d$ simulations result in increased horizontal shear.  Our findings suggest that large scale moist convection may be the dominant cause of the formation of high latitude FFRs, which is in support of previous observations \citep[e.g.][]{Borucki1992}.  Furthermore, we find that high $L_d$ simulations show Arnol'd-stable behavior using the vortical Mach number condition, ``Ma"$^{-1}=1$ \citep{Dowling2020}.  Higher deformation lengths are able to stabilize the domain via homogenization of local potential vorticity in the form of polar cyclone formation.  Low $L_d$ simulations are difficult to interpret with regards to Arnol'd-stability, but show a clear distinction from the high $L_d$ simulations.

%{\bf{We also note the formation of dominant features with wavenumber dependence on the deformation lengths.  Our approach to the changing zonal wavenumbers is purely geometric.  Increasing the deformation length (decreasing planetary Bu) reduces the number of balanced eddies that stabilize at the polar regions.  A more detailed treatment should also look at the dominant frequencies that are active for a given wave number.  Morphologically, this behavior is reminiscent of polar vortex Rossby waves \citep{Chen&Yau2001} and should be studied further.}} 

We also note that the emergence of patterns with different dominant wavenumbers in our simulations depends primarily on the size (relative to that of our domain) of the largest balanced vortex that can emerge in each simulation. Increasing the deformation length (decreasing planetary Bu) reduces the number of balanced eddies that stabilize at the polar regions.  The wavenumber that emerges is dependent on number of the balanced eddies, as well as how they self-organize within the domain.  Morphologically, this behavior is reminiscent of polar vortex Rossby waves \citep{Chen&Yau2001, Houghton2002} and should be studied further. 

Our work can be expanded upon with the inclusion of stratification.  The model can be made multilayered to include baroclinicity and thermodynamical effects.  The parameter space may be reduced to study low deformation length cases at significantly higher resolution while forcing the domain using baroclinic modes.  As multilayered shallow water systems crudely approximate stratification, allowing baroclinic modes in a multilayered system will allow a better quantification of the effect of anisotropic turbulence on the upper atmosphere and how it modulates the long term behavior of these flows.  Further modifications may include variations in the underlying dynamical topography.  For simplicity, we use no variation in the underlying geopotential for a given simulation.  Added complexities may lead to further dynamically interesting behavior as it may impede meridional eddy momentum flux transfer over long timescales.

% Although this is beyond the scope of our work, our model can be used to investigate these emergent features in context of vortex Rossby waves \citep{Chen&Yau2001}.}}

% Although it is beyond the scope of this work, this should be investigated further in the context of vortex Rossby waves, which exhibit morphologies similar to our simulations \citep{Chen&Yau2001}.}}

\acknowledgments

We thank two anonymous referees for their useful comments that helped improve our manuscript.  We also thank Dr. John Rogers and Gerald Eichst{\"a}dt for the contribution of the JunoCam image.  A.H. also thanks Dr. Shawn Brueshaber and Dr. Morgan O'Neill for important discussions regarding gas giant polar modeling and providing important insight into the parameter space.  This work was supported by NASA’s New Frontiers Data Analysis Program (grant\# 80NSSC20K0561) and by the NASA Fellowship Activity (grant\# 80NSSC20K1456).  The authors acknowledge the Texas Advanced Computing Center (TACC) at The University of Texas at Austin for providing HPC resources that have contributed to the research results reported within this paper (\url{http://www.tacc.utexas.edu}).

\software{{\sc{The Pencil Code}} \citep{Brandenburg+2002, BrandenburgDobler10, Brandenburg+Scannapieco2020},
           Numpy \citep{numpy2011, numpy2020},  Scipy \citep{scipy2001},  Matplotlib \citep{Hunter2007}
          }

\appendix

\section{Scaling Study} \label{sec:app_scaling}

\begin{figure}[h]
    \centering
    \includegraphics[width=0.45\textwidth]{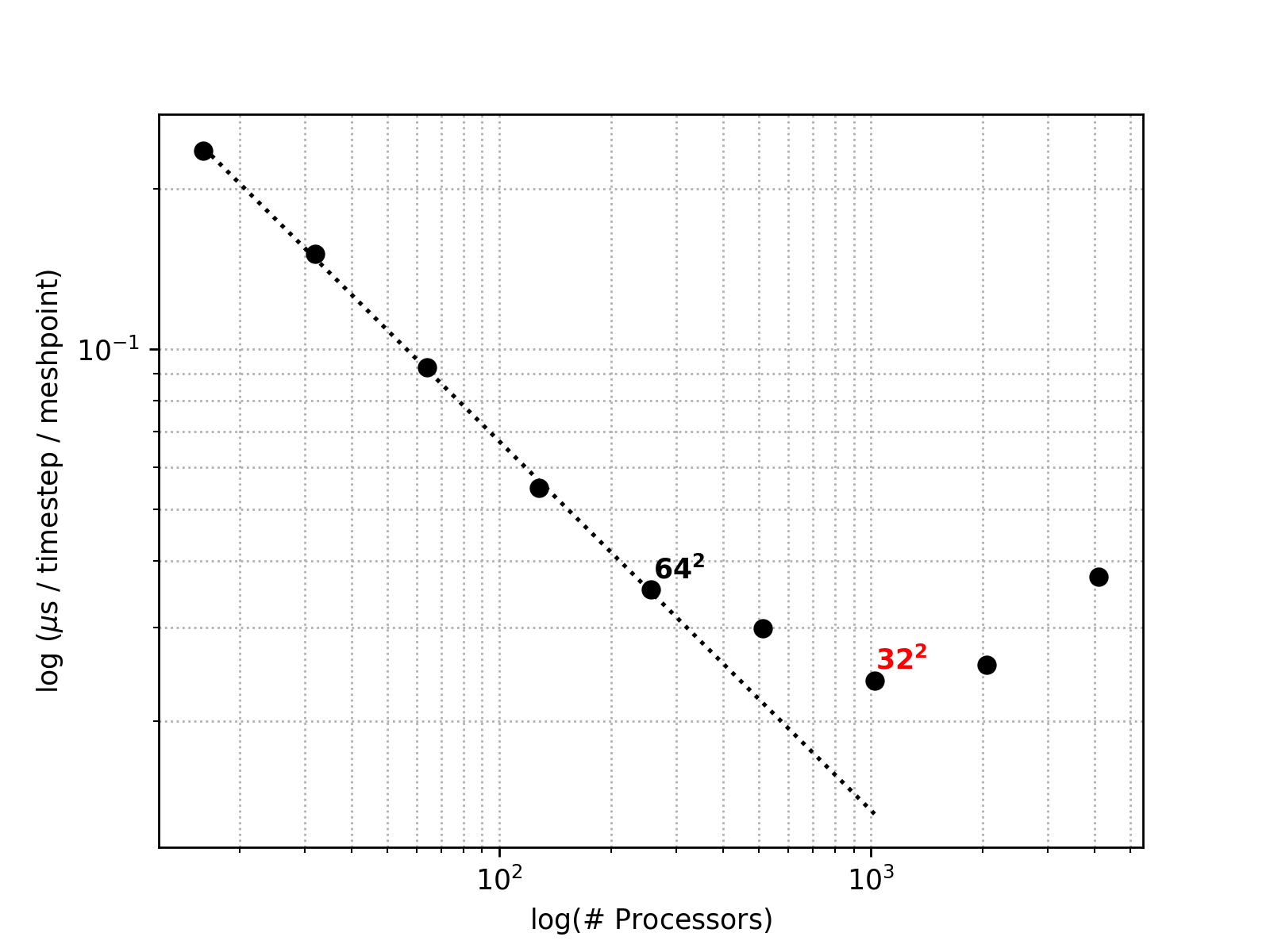}
    \caption{Strong scaling study using Stampede2 at a base resolution of $1024^2$.  With our fixed resolution, $32^2$ optimizes the balance between computation and communication time.}
    \label{f:strong}
\end{figure}

As we are presenting a new shallow water atmospheric module in Pencil, it is important to assess the efficiency of our code using a scaling analysis that characterizes the effectiveness of parallelization.  We employed the XSEDE/Stampede2 supercomputer at Texas Advanced Computing Center (TACC) for our simulations.  We perform a strong scaling analysis to determine how the code handles large scale communication between active nodes on the cluster.  We fix the resolution at $1024^2$ while keeping all of the parameters for a full simulation active, ensuring that the timestep average would be representative of simulations presented in this paper.  In Figure \ref{f:strong}, we see that a processor load of $32^2$ grid points minimizes total computation and communication most effectively.  This load signifies the turnover point where communication time is already causing computational inefficiency, however, we elect to use it as the increase in time from processor loads of $64^2$ or $32\times64$ is still important for simulations with long term evolution.

\begin{figure}[h]
    \centering
    \includegraphics[width=0.45\textwidth]{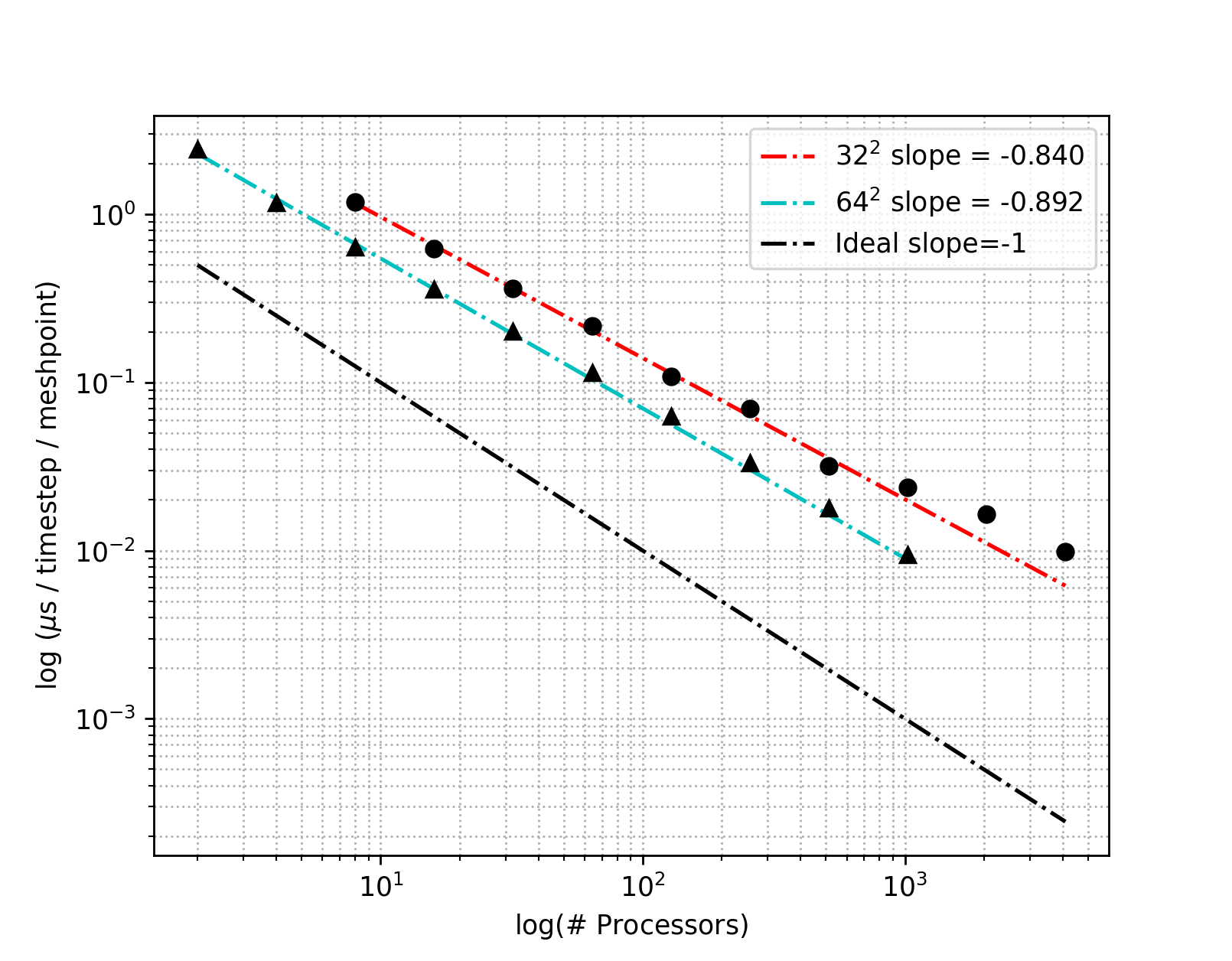}
    \caption{Weak scaling study using Stampede2 with varying processing loads.  The red dash-dotted line is for a processor load of $32^2$ while the cyan is for $64^2$.}
    \label{f:weak}
\end{figure}

The efficiency of the parallelization in Pencil using weak scaling is shown in Figure \ref{f:weak}.  As expected, maintaining the optimal processor load, $32^2$, and increasing the resolution via increasing total computing nodes used yields a highly linear relation, indicative of efficient parallelization.  Tests were conducted up to resolutions of $2048^2$; at a resolution of $1024^2$, the $32^2$ processor load optimizes the simulations over $64^2$.  Thus, our simulations can be used for studying the polar dynamics with $\times16$ more resolution than previous simulations if a resolution of $2048^2$ is employed.

\section{Benchmark} \label{sec:app_Benchmarks}

Since our model uses approximations and methodologies explored in previous works, we opt to test the validity of our model using a phenomenological benchmark.  We use the vortex solution developed by \citet{DeMaria1984}.  This solution has been utilized in finite-volume methods to address the stability of the Jovian polar vortex configurations and merger events \citep{Li+2020}.  Here, we use it as a measure of the stability of our code.  The radial velocity profile, $v(r)$, for the vortex can be written as:
\begin{equation}
\begin{split}
v(r) = v_m \left(\frac{r}{r_m}\right) \exp{\left\{\frac{1}{b}\left[1-\left(\frac{r}{r_m}\right)^b\right]\right\}},
\end{split}
\label{e:app_demaria-vortex}
\end{equation}

\noindent where $v_m$ is the maximum tangential velocity, $r$ is the distance from the center of the vortex, $r_m$ is the radius of maximum wind speed, and $b$ is a shielding parameter, which determines the strength of the surrounding (opposite-signed) vorticity.  Equation \ref{e:app_demaria-vortex} is a numerical approximation of the behavior of a tropical cyclone interacting with another cyclone in its vicinity.  A full conceptual argument of this interaction is provided in \citep{Chang1983} and \citep{DeMaria1984}.  Likewise, the corresponding geopotential profile is calculated numerically.  This results in a minor phase of adjustment during which the geopotential and the velocity profile must relax into their equilibrium values.  The adjustment phase only lasts for a brief period and imparts gravity waves until the velocity field and geopotential appropriately converge.

\begin{figure*}[hbtp] \center
    \includegraphics[width=\textwidth]{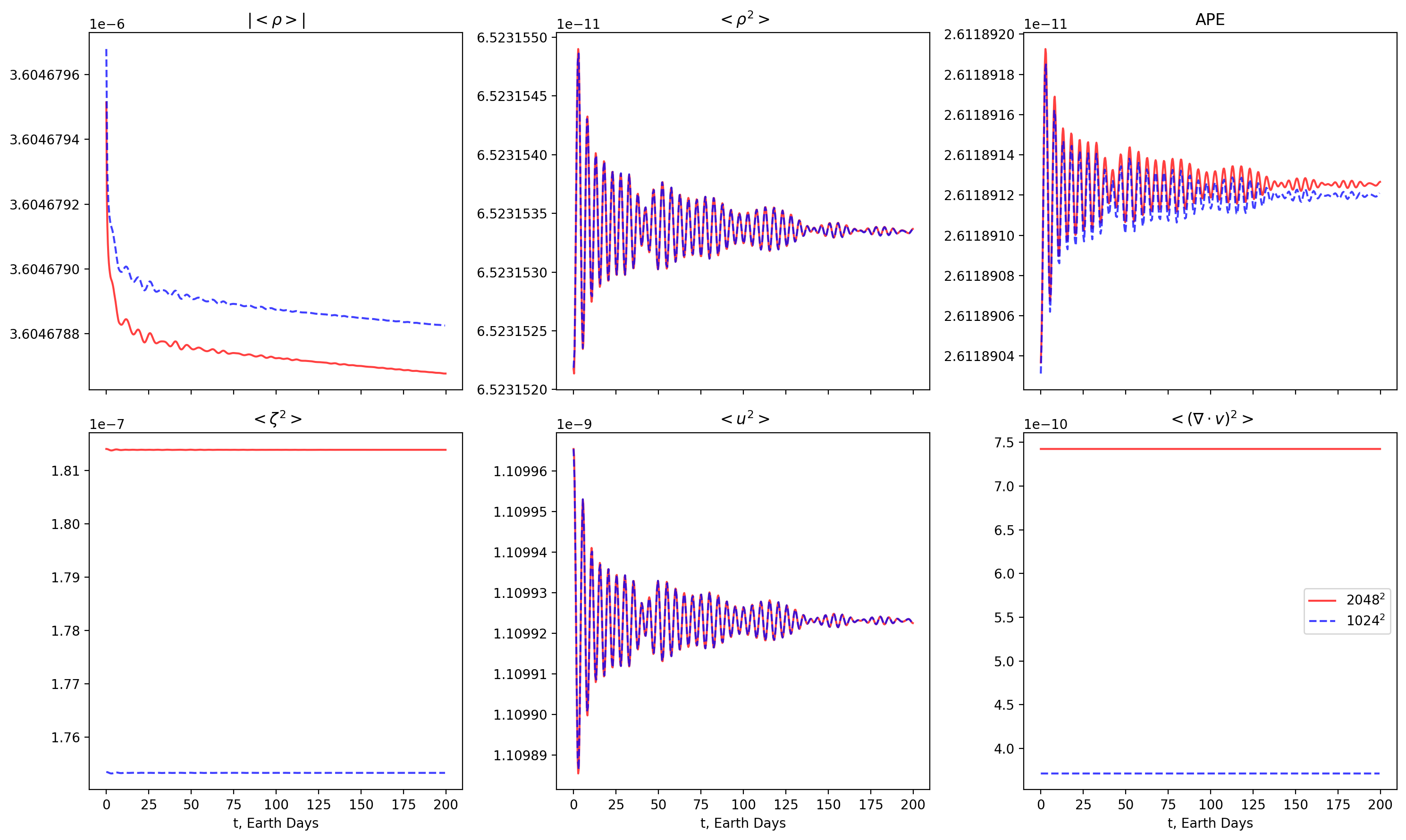}
    \caption{Behavior of diagnostic quantities of interest given the initial vortex solution of \citet{Li+2020, DeMaria1984}.  The values stabilize by about $200$ Earth days and show sufficient convergence from resolutions $1024^2$ (dashed blue) to $2048^2$ (solid red).}
    \label{f:diagnostics}
\end{figure*}

In Figure \ref{f:diagnostics}, we show the early evolution of the vortex solution from \citep{DeMaria1984}.  We offset the vortex from the pole by a small amount to highlight any azimuthal asymmetries that may exist in the simulation.  The vortex stabilizes at about $200$ Earth days and varies very slowly and no azimuthal asymmetries are seen.  To test convergence, the model is run at resolutions $1024^2$ and $2048^2$.  All diagnostics show code convergence as the differences are only on the order of $10^{-6}$.  The maximum relative error is exhibited in the mean-square divergence.  However, as we employ incompressibility of the flow in question, the overall effect is on the order of $10^{-10}$.  Both KE (in the form of $\langle u^2\rangle$) and APE tend to oscillate and converge in the same manner for both resolutions, and are sufficiently stable by day $200$.  The adjustment phase is dependent upon the numerical integration scheme used for the initial geopotential as larger errors will impart stronger gravity waves as the flow adjusts to the initialized velocity profile given by equation \ref{e:app_demaria-vortex}.  However, low-order numerical integration only effects low resolutions and is not a concern for the resolutions used in this work.  Thus, employing $1024^2$ for our simulations is sufficient to capture relevant dynamical behavior of a $1\frac{1}{2}$-layer shallow water system.

%% For this sample we use BibTeX plus aasjournals.bst to generate the
%% the bibliography. The sample63.bib file was populated from ADS. To
%% get the citations to show in the compiled file do the following:
%%
%% pdflatex sample63.tex
%% bibtext sample63
%% pdflatex sample63.tex
%% pdflatex sample63.tex

\bibliography{bibtex}{}

\begin{thebibliography}{}
\expandafter\ifx\csname natexlab\endcsname\relax\def\natexlab#1{#1}\fi
\providecommand{\url}[1]{\href{#1}{#1}}
\providecommand{\dodoi}[1]{doi:~\href{http://doi.org/#1}{\nolinkurl{#1}}}
\providecommand{\doeprint}[1]{\href{http://ascl.net/#1}{\nolinkurl{http://ascl.net/#1}}}
\providecommand{\doarXiv}[1]{\href{https://arxiv.org/abs/#1}{\nolinkurl{https://arxiv.org/abs/#1}}}

\bibitem[{{Adriani} {et~al.}(2017){Adriani}, {Filacchione}, {Di Iorio},
  {Turrini}, {Noschese}, {Cicchetti}, {Grassi}, {Mura}, {Sindoni}, {Zambelli},
  {Piccioni}, {Capria}, {Tosi}, {Orosei}, {Dinelli}, {Moriconi}, {Roncon},
  {Lunine}, {Becker}, {Bini}, {Barbis}, {Calamai}, {Pasqui}, {Nencioni},
  {Rossi}, {Lastri}, {Formaro}, \& {Olivieri}}]{jiram}
{Adriani}, A., {Filacchione}, G., {Di Iorio}, T., {et~al.} 2017, Space Science
  Reviews, 213, 393, \dodoi{10.1007/s11214-014-0094-y}

\bibitem[{{Adriani} {et~al.}(2018){Adriani}, {Mura}, {Orton}, {Hansen},
  {Altieri}, {Moriconi}, {Rogers}, {Eichst{\"a}dt}, {Momary}, {Ingersoll},
  {Filacchione}, {Sindoni}, {Tabataba-Vakili}, {Dinelli}, {Fabiano}, {Bolton},
  {Connerney}, {Atreya}, {Lunine}, {Tosi}, {Migliorini}, {Grassi}, {Piccioni},
  {Noschese}, {Cicchetti}, {Plainaki}, {Olivieri}, {O'Neill}, {Turrini},
  {Stefani}, {Sordini}, \& {Amoroso}}]{Adriani+2018}
{Adriani}, A., {Mura}, A., {Orton}, G., {et~al.} 2018, \nat, 555, 216,
  \dodoi{10.1038/nature25491}

\bibitem[{{Andrews}(1984)}]{Andrews1984}
{Andrews}, D.~G. 1984, Geophysical and Astrophysical Fluid Dynamics, 28, 243,
  \dodoi{10.1080/03091928408230366}

\bibitem[{{Bagenal} {et~al.}(2007){Bagenal}, {Dowling}, \&
  {McKinnon}}]{jupiter2007}
{Bagenal}, F., {Dowling}, T.~E., \& {McKinnon}, W.~B. 2007, {Jupiter}

\bibitem[{{Bardet} {et~al.}(2019){Bardet}, {Spiga}, {Guerlet}, {Millour},
  {Boissinot}, \& {Dubos}}]{DYNAMICO_b}
{Bardet}, D., {Spiga}, A., {Guerlet}, S., {et~al.} 2019, in EGU General
  Assembly Conference Abstracts, EGU General Assembly Conference Abstracts,
  14516

\bibitem[{{Boissinot} {et~al.}(2019){Boissinot}, {Spiga}, {Guerlet}, {Cabanes},
  \& {Hourdin}}]{DYNAMICO_a}
{Boissinot}, A., {Spiga}, A., {Guerlet}, S., {Cabanes}, S., \& {Hourdin}, F.
  2019, in EGU General Assembly Conference Abstracts, EGU General Assembly
  Conference Abstracts, 15821

\bibitem[{{Borucki} \& {Magalhaes}(1992)}]{Borucki1992}
{Borucki}, W.~J., \& {Magalhaes}, J.~A. 1992, \icarus, 96, 1,
  \dodoi{10.1016/0019-1035(92)90002-O}

\bibitem[{{Bourdin} {et~al.}(2013){Bourdin}, {Bingert}, \&
  {Peter}}]{Bourdin+2013}
{Bourdin}, P.~A., {Bingert}, S., \& {Peter}, H. 2013, \aap, 555, A123,
  \dodoi{10.1051/0004-6361/201321185}

\bibitem[{{Brandenburg} \& {Dobler}(2002)}]{Brandenburg+2002}
{Brandenburg}, A., \& {Dobler}, W. 2002, Computer Physics Communications, 147,
  471, \dodoi{10.1016/S0010-4655(02)00334-X}

\bibitem[{{Brandenburg} \& {Dobler}(2010)}]{BrandenburgDobler10}
---. 2010, {Pencil: Finite-difference Code for Compressible Hydrodynamic
  Flows}, Astrophysics Source Code Library.
\newblock \doeprint{1010.060}

\bibitem[{{Brandenburg} \& {Scannapieco}(2020)}]{Brandenburg+Scannapieco2020}
{Brandenburg}, A., \& {Scannapieco}, E. 2020, \apj, 889, 55,
  \dodoi{10.3847/1538-4357/ab5e7f}

\bibitem[{{Bridger} \& {Stevens}({1980})}]{Bridger&Stevens1980}
{Bridger}, A. F.~C., \& {Stevens}, D.~E. {1980}, Journal of Atmospheric
  Sciences, 37, 534 , \dodoi{10.1175/1520-0469(1980)037<0534:LAWATP>2.0.CO;2}

\bibitem[{{Brueshaber} \& {Sayanagi}(2021)}]{Brueshaber+2021}
{Brueshaber}, S.~R., \& {Sayanagi}, K.~M. 2021, \icarus, 361, 114386,
  \dodoi{10.1016/j.icarus.2021.114386}

\bibitem[{{Brueshaber} {et~al.}(2019){Brueshaber}, {Sayanagi}, \&
  {Dowling}}]{Brueshaber+2019}
{Brueshaber}, S.~R., {Sayanagi}, K.~M., \& {Dowling}, T.~E. 2019, \icarus, 323,
  46, \dodoi{10.1016/j.icarus.2019.02.001}

\bibitem[{Cai {et~al.}(2021)Cai, Chan, \& Mayr}]{Cai2021DeepCP}
Cai, T., Chan, K.~L., \& Mayr, H.~G. 2021, The Planetary Science Journal, 2

\bibitem[{{Carnevale} \& {Shepherd}(1990)}]{Carnevale+1990}
{Carnevale}, G.~F., \& {Shepherd}, T.~G. 1990, Geophysical and Astrophysical
  Fluid Dynamics, 51, 1, \dodoi{10.1080/03091929008219847}

\bibitem[{{Chandrasekhar}(1961)}]{Chandrasekhar1961}
{Chandrasekhar}, S. 1961, {Hydrodynamic and hydromagnetic stability}

\bibitem[{{Chang}(1983)}]{Chang1983}
{Chang}, S. W.-J. 1983, Monthly Weather Review, 111, 1806 ,
  \dodoi{10.1175/1520-0493(1983)111<1806:ANSOTI>2.0.CO;2}

\bibitem[{Chen \& Yau(2001)}]{Chen&Yau2001}
Chen, Y., \& Yau, M.~K. 2001, Journal of the Atmospheric Sciences, 58, 2128 ,
  \dodoi{10.1175/1520-0469(2001)058<2128:SBIASH>2.0.CO;2}

\bibitem[{{DeMaria} \& {Chan}(1984)}]{DeMaria1984}
{DeMaria}, M., \& {Chan}, J. C.~L. 1984, Monthly Weather Review, 112, 1643 ,
  \dodoi{10.1175/1520-0493(1984)112<1643:CONSOT>2.0.CO;2}

\bibitem[{{Dowling}(1993)}]{Dowling1993}
{Dowling}, T.~E. 1993, Journal of Atmospheric Sciences, 50, 14 ,
  \dodoi{10.1175/1520-0469(1993)050<0014:ARBPVA>2.0.CO;2}

\bibitem[{{Dowling}(2020)}]{Dowling2020}
---. 2020, The Planetary Science Journal, 1, 6, \dodoi{10.3847/PSJ/ab789d}

\bibitem[{{Dowling} {et~al.}(1998){Dowling}, {Fischer}, {Gierasch},
  {Harrington}, {LeBeau}, \& {Santori}}]{Dowling+1998}
{Dowling}, T.~E., {Fischer}, A.~S., {Gierasch}, P.~J., {et~al.} 1998, \icarus,
  132, 221, \dodoi{10.1006/icar.1998.5917}

\bibitem[{Dowling \& Ingersoll(1989)}]{Dowling&Ingersoll1989}
Dowling, T.~E., \& Ingersoll, A.~P. 1989, Journal of Atmospheric Sciences, 46,
  3256 , \dodoi{10.1175/1520-0469(1989)046<3256:JGRSAA>2.0.CO;2}

\bibitem[{{Fletcher} {et~al.}(2017){Fletcher}, {Orton}, {Rogers}, {Giles},
  {Payne}, {Irwin}, \& {Vedovato}}]{Fletcher2017}
{Fletcher}, L.~N., {Orton}, G.~S., {Rogers}, J.~H., {et~al.} 2017, \icarus,
  286, 94, \dodoi{10.1016/j.icarus.2017.01.001}

\bibitem[{Fodor(2003)}]{Fodor2003}
Fodor, F. 2003, Beiträge zur Algebra und Geometrie, 44

\bibitem[{{Garcia} {et~al.}(2020){Garcia}, {Chambers}, \&
  {Watts}}]{Garcia+2020}
{Garcia}, F., {Chambers}, F. R.~N., \& {Watts}, A.~L. 2020, \mnras, 499, 4698,
  \dodoi{10.1093/mnras/staa2962}

\bibitem[{{Garc{\'\i}a-Melendo} {et~al.}(2013){Garc{\'\i}a-Melendo}, {Hueso},
  {S{\'a}nchez-Lavega}, {Legarreta}, {Del R{\'\i}o-Gaztelurrutia},
  {P{\'e}rez-Hoyos}, \& {Sanz-Requena}}]{GarciaMelendo2013}
{Garc{\'\i}a-Melendo}, E., {Hueso}, R., {S{\'a}nchez-Lavega}, A., {et~al.}
  2013, Nature Geoscience, 6, 525, \dodoi{10.1038/ngeo1860}

\bibitem[{{García-Melendo} \& {Sánchez-Lavega}(2017)}]{GARCIAMELENDO2017}
{García-Melendo}, E., \& {Sánchez-Lavega}, A. 2017, Icarus, 286, 241,
  \dodoi{https://doi.org/10.1016/j.icarus.2016.10.006}

\bibitem[{{Gavriel} \& {Kaspi}(2021)}]{Gavriel2021}
{Gavriel}, N., \& {Kaspi}, Y. 2021, Nature Geoscience, 14, 559,
  \dodoi{10.1038/s41561-021-00781-6}

\bibitem[{{Godfrey}(1988)}]{Godfrey1998}
{Godfrey}, D.~A. 1988, \icarus, 76, 335, \dodoi{10.1016/0019-1035(88)90075-9}

\bibitem[{Golub \& Van~Loan(1996)}]{Golub}
Golub, G.~H., \& Van~Loan, C.~F. 1996, Matrix Computations (3rd Ed.) (USA:
  Johns Hopkins University Press)

\bibitem[{{Grassi} {et~al.}(2018){Grassi}, {Adriani}, {Moriconi}, {Mura},
  {Tabataba-Vakili}, {Ingersoll}, {Orton}, {Hansen}, {Altieri}, {Filacchione},
  {Sindoni}, {Dinelli}, {Fabiano}, {Bolton}, {Levin}, {Atreya}, {Lunine},
  {Momary}, {Tosi}, {Migliorini}, {Piccioni}, {Noschese}, {Cicchetti},
  {Plainaki}, {Olivieri}, {Turrini}, {Stefani}, {Sordini}, \&
  {Amoroso}}]{Grassi+2018}
{Grassi}, D., {Adriani}, A., {Moriconi}, M.~L., {et~al.} 2018, Journal of
  Geophysical Research (Planets), 123, 1511, \dodoi{10.1029/2018JE005555}

\bibitem[{{Harris} {et~al.}(2020){Harris}, {Millman}, {van der Walt},
  {Gommers}, {Virtanen}, {Cournapeau}, {Wieser}, {Taylor}, {Berg}, {Smith},
  {Kern}, {Picus}, {Hoyer}, {van Kerkwijk}, {Brett}, {Haldane}, {del R{\'\i}o},
  {Wiebe}, {Peterson}, {G{\'e}rard-Marchant}, {Sheppard}, {Reddy}, {Weckesser},
  {Abbasi}, {Gohlke}, \& {Oliphant}}]{numpy2020}
{Harris}, C.~R., {Millman}, K.~J., {van der Walt}, S.~J., {et~al.} 2020, \nat,
  585, 357, \dodoi{10.1038/s41586-020-2649-2}

\bibitem[{{Haugen} {et~al.}(2004{\natexlab{a}}){Haugen}, {Brandenburg}, \&
  {Dobler}}]{Haugen+2004}
{Haugen}, N.~E., {Brandenburg}, A., \& {Dobler}, W. 2004{\natexlab{a}}, \pre,
  70, 016308, \dodoi{10.1103/PhysRevE.70.016308}

\bibitem[{{Haugen} {et~al.}(2004{\natexlab{b}}){Haugen}, {Brandenburg}, \&
  {Mee}}]{Haugen++2004}
{Haugen}, N. E.~L., {Brandenburg}, A., \& {Mee}, A.~J. 2004{\natexlab{b}},
  \mnras, 353, 947, \dodoi{10.1111/j.1365-2966.2004.08127.x}

\bibitem[{Hogg \& Stommel(1985)}]{hetons_1985}
Hogg, N.~G., \& Stommel, H.~M. 1985, Proceedings of the Royal Society of
  London. A. Mathematical and Physical Sciences, 397, 1,
  \dodoi{10.1098/rspa.1985.0001}

\bibitem[{{Hord} {et~al.}(2017){Hord}, {Lyra}, {Flock}, {Turner}, \& {Mac
  Low}}]{Hord+17}
{Hord}, B., {Lyra}, W., {Flock}, M., {Turner}, N.~J., \& {Mac Low}, M.-M. 2017,
  \apj, 849, 164, \dodoi{10.3847/1538-4357/aa8fcf}

\bibitem[{{Houghton}(2002)}]{Houghton2002}
{Houghton}, J. 2002, {The Physics of Atmospheres}

\bibitem[{Hunter(2007)}]{Hunter2007}
Hunter, J.~D. 2007, Computing in Science \& Engineering, 9, 90,
  \dodoi{10.1109/MCSE.2007.55}

\bibitem[{Hyder {et~al.}(2022)Hyder, Lyra, Chanover, Morales-Juberías, \&
  Jackiewicz}]{hyder_ali_2022_6642986}
Hyder, A., Lyra, W., Chanover, N., Morales-Juberías, R., \& Jackiewicz, J.
  2022, {Exploring Jupiter's Polar Deformation Lengths with High Resolution
  Shallow Water Modeling},  Zenodo, \dodoi{10.5281/zenodo.6642986}

\bibitem[{{Jin} \& {Dubin}(1998)}]{JinDubin1998}
{Jin}, D.~Z., \& {Dubin}, D. H.~E. 1998, \prl, 80, 4434,
  \dodoi{10.1103/PhysRevLett.80.4434}

\bibitem[{Jones {et~al.}(2001)Jones, Oliphant, \& Peterson}]{scipy2001}
Jones, E., Oliphant, T., \& Peterson, P. 2001

\bibitem[{{Kaspi}(2008)}]{Kaspi2008PhD}
{Kaspi}, Y. 2008, PhD thesis, Massachusetts Institute of Technology

\bibitem[{{Kaspi} {et~al.}(2009){Kaspi}, {Flierl}, \& {Showman}}]{Kaspi+2009}
{Kaspi}, Y., {Flierl}, G.~R., \& {Showman}, A.~P. 2009, \icarus, 202, 525,
  \dodoi{10.1016/j.icarus.2009.03.026}

\bibitem[{Klemp {et~al.}(2018)Klemp, Skamarock, \& Ha}]{Klemp+2018}
Klemp, J.~B., Skamarock, W.~C., \& Ha, S. 2018, Monthly Weather Review, 146,
  1911 , \dodoi{10.1175/MWR-D-17-0384.1}

\bibitem[{{Li} {et~al.}(2020){Li}, {Ingersoll}, {Klipfel}, \&
  {Brettle}}]{Li+2020}
{Li}, C., {Ingersoll}, Andrew, P., {Klipfel}, Alexandra, P., \& {Brettle}, H.
  2020, Proceedings of the National Academy of Science, 117, 24082,
  \dodoi{10.1073/pnas.2008440117}

\bibitem[{{Lyra} {et~al.}(2008){Lyra}, {Johansen}, {Klahr}, \&
  {Piskunov}}]{Lyra+08a}
{Lyra}, W., {Johansen}, A., {Klahr}, H., \& {Piskunov}, N. 2008, \aap, 479,
  883, \dodoi{10.1051/0004-6361:20077948}

\bibitem[{{Lyra} {et~al.}(2017){Lyra}, {McNally}, {Heinemann}, \&
  {Masset}}]{Lyra+17}
{Lyra}, W., {McNally}, C.~P., {Heinemann}, T., \& {Masset}, F. 2017, \aj, 154,
  146, \dodoi{10.3847/1538-3881/aa8811}

\bibitem[{{Lyra} {et~al.}(2018){Lyra}, {Raettig}, \& {Klahr}}]{Lyra+18}
{Lyra}, W., {Raettig}, N., \& {Klahr}, H. 2018, Research Notes of the American
  Astronomical Society, 2, 195, \dodoi{10.3847/2515-5172/aaeac9}

\bibitem[{{Lyra} {et~al.}(2016){Lyra}, {Richert}, {Boley}, {Turner}, {Mac Low},
  {Okuzumi}, \& {Flock}}]{Lyra+16}
{Lyra}, W., {Richert}, A. J.~W., {Boley}, A., {et~al.} 2016, \apj, 817, 102,
  \dodoi{10.3847/0004-637X/817/2/102}

\bibitem[{{Lyra} {et~al.}(2015){Lyra}, {Turner}, \& {McNally}}]{Lyra+15}
{Lyra}, W., {Turner}, N.~J., \& {McNally}, C.~P. 2015, \aap, 574, A10,
  \dodoi{10.1051/0004-6361/201424919}

\bibitem[{{Marcus}(2004)}]{Marcus2004}
{Marcus}, P.~S. 2004, \nat, 428, 828, \dodoi{10.1038/nature02470}

\bibitem[{Marshall(2008)}]{MarshallJohn2008Aoac}
Marshall, J. 2008, Atmosphere, ocean, and climate dynamics an introductory
  text, International geophysics series ; v. 93 (Amsterdam ; [Burlington, MA]:
  Elsevier Academic Press)

\bibitem[{Marshall {et~al.}(1997)Marshall, Adcroft, Hill, Perelman, \&
  Heisey}]{MITgcm}
Marshall, J., Adcroft, A., Hill, C., Perelman, L., \& Heisey, C. 1997, Journal
  of Geophysical Research: Oceans, 102, 5753,
  \dodoi{https://doi.org/10.1029/96JC02775}

\bibitem[{Mcintyre \& Shepherd(1987)}]{McintyreShepherd1987}
Mcintyre, M., \& Shepherd, T. 1987, Journal of Fluid Mechanics, 181, 527 ,
  \dodoi{10.1017/S0022112087002209}

\bibitem[{{Montgomery} {et~al.}(2006){Montgomery}, {Nicholls}, {Cram}, \&
  {Saunders}}]{Montgomery2006}
{Montgomery}, M.~T., {Nicholls}, M.~E., {Cram}, T.~A., \& {Saunders}, A.~B.
  2006, Journal of Atmospheric Sciences, 63, 355, \dodoi{10.1175/JAS3604.1}

\bibitem[{{Morales-Juber{\'\i}as} {et~al.}(2011){Morales-Juber{\'\i}as},
  {Sayanagi}, {Dowling}, \& {Ingersoll}}]{Morales-Juberias+2011}
{Morales-Juber{\'\i}as}, R., {Sayanagi}, K.~M., {Dowling}, T.~E., \&
  {Ingersoll}, A.~P. 2011, \icarus, 211, 1284,
  \dodoi{10.1016/j.icarus.2010.11.006}

\bibitem[{{Morales-Juber{\'\i}as} {et~al.}(2015){Morales-Juber{\'\i}as},
  {Sayanagi}, {Simon}, {Fletcher}, \& {Cosentino}}]{Morales-Juberias+2015}
{Morales-Juber{\'\i}as}, R., {Sayanagi}, K.~M., {Simon}, A.~A., {Fletcher},
  L.~N., \& {Cosentino}, R.~G. 2015, \apjl, 806, L18,
  \dodoi{10.1088/2041-8205/806/1/L18}

\bibitem[{Mu \& Wu(2001)}]{MuWu2001}
Mu, M., \& Wu, Y.-H. 2001, Surveys in Geophysics, 22, 383,
  \dodoi{10.1023/A:1014229917728}

\bibitem[{{O'Neill} {et~al.}(2015){O'Neill}, {Emanuel}, \&
  {Flierl}}]{ONeil+2015}
{O'Neill}, M.~E., {Emanuel}, K.~A., \& {Flierl}, G.~R. 2015, Nature Geoscience,
  8, 523, \dodoi{10.1038/ngeo2459}

\bibitem[{{Orton} {et~al.}(2017){Orton}, {Hansen}, {Caplinger}, {Ravine},
  {Atreya}, {Ingersoll}, {Jensen}, {Momary}, {Lipkaman}, {Krysak}, {Zimdar}, \&
  {Bolton}}]{Orton+2017}
{Orton}, G.~S., {Hansen}, C., {Caplinger}, M., {et~al.} 2017, \grl, 44, 4599,
  \dodoi{10.1002/2016GL072443}

\bibitem[{{O’Neill} {et~al.}(2016){O’Neill}, Emanuel, \&
  Flierl}]{ONeill+2016}
{O’Neill}, M.~E., Emanuel, K.~A., \& Flierl, G.~R. 2016, Journal of the
  Atmospheric Sciences, 73, 1841 , \dodoi{10.1175/JAS-D-15-0314.1}

\bibitem[{Pedlosky(1979)}]{Pedlosky1979}
Pedlosky, J. 1979, Geophysical fluid dynamics / Joseph Pedlosky (Springer
  Verlag New York), xii, 624 p. :

\bibitem[{{Pedlosky}(1985)}]{Pedlosky1085}
{Pedlosky}, J. 1985, Journal of Atmospheric Sciences, 42, 1477 ,
  \dodoi{10.1175/1520-0469(1985)042<1477:TIOCHC>2.0.CO;2}

\bibitem[{Pirl(1969)}]{Pirl1969DerMV}
Pirl, U. 1969, Mathematische Nachrichten, 40, 111

\bibitem[{{Raettig} {et~al.}(2015){Raettig}, {Klahr}, \& {Lyra}}]{Raettig+2015}
{Raettig}, N., {Klahr}, H., \& {Lyra}, W. 2015, \apj, 804, 35,
  \dodoi{10.1088/0004-637X/804/1/35}

\bibitem[{{Raettig} {et~al.}(2013){Raettig}, {Lyra}, \& {Klahr}}]{Raettig+13}
{Raettig}, N., {Lyra}, W., \& {Klahr}, H. 2013, \apj, 765, 115,
  \dodoi{10.1088/0004-637X/765/2/115}

\bibitem[{{Read} {et~al.}(2020){Read}, {Kennedy}, {Lewis}, {Scolan},
  {Tabataba-Vakili}, {Wang}, {Wright}, \& {Young}}]{Read2020}
{Read}, P., {Kennedy}, D., {Lewis}, N., {et~al.} 2020, Nonlinear Processes in
  Geophysics, 27, 147, \dodoi{10.5194/npg-27-147-2020}

\bibitem[{{Read} {et~al.}(2006){Read}, {Gierasch}, {Conrath}, {Simon-Miller},
  {Fouchet}, \& {Yamazaki}}]{Read+2006a}
{Read}, P.~L., {Gierasch}, P.~J., {Conrath}, B.~J., {et~al.} 2006, Quarterly
  Journal of the Royal Meteorological Society, 132, 1577,
  \dodoi{10.1256/qj.05.34}

\bibitem[{{Richert} {et~al.}(2015){Richert}, {Lyra}, {Boley}, {Mac Low}, \&
  {Turner}}]{Richert+15}
{Richert}, A. J.~W., {Lyra}, W., {Boley}, A., {Mac Low}, M.-M., \& {Turner}, N.
  2015, \apj, 804, 95, \dodoi{10.1088/0004-637X/804/2/95}

\bibitem[{Rogers {et~al.}(2022)Rogers, Eichstädt, Hansen, Orton, Momary,
  Casely, Adamoli, Jacquesson, Bullen, Peach, Olivetti, Brueshaber, Ravine, \&
  Bolton}]{ROGERS2022114742}
Rogers, J., Eichstädt, G., Hansen, C., {et~al.} 2022, Icarus, 372, 114742,
  \dodoi{https://doi.org/10.1016/j.icarus.2021.114742}

\bibitem[{{S{\'a}nchez-Lavega} {et~al.}(2011){S{\'a}nchez-Lavega}, {del
  R{\'\i}o-Gaztelurrutia}, {Hueso}, {G{\'o}mez-Forrellad}, {Sanz-Requena},
  {Legarreta}, {Garc{\'\i}a-Melendo}, {Colas}, {Lecacheux}, {Fletcher},
  {Barrado y Navascu{\'e}s}, {Parker}, {International Outer Planet Watch Team},
  {Akutsu}, {Barry}, {Beltran}, {Buda}, {Combs}, {Carvalho}, {Casquinha},
  {Delcroix}, {Ghomizadeh}, {Go}, {Hotershall}, {Ikemura}, {Jolly}, {Kazemoto},
  {Kumamori}, {Lecompte}, {Maxson}, {Melillo}, {Milika}, {Morales}, {Peach},
  {Phillips}, {Poupeau}, {Sussenbach}, {Walker}, {Walker}, {Tranter}, {Wesley},
  {Wilson}, \& {Yunoki}}]{SanchezLavega2011}
{S{\'a}nchez-Lavega}, A., {del R{\'\i}o-Gaztelurrutia}, T., {Hueso}, R.,
  {et~al.} 2011, \nat, 475, 71, \dodoi{10.1038/nature10203}

\bibitem[{Scott(2011)}]{Scott}
Scott, R. 2011, Geophysical \& Astrophysical Fluid Dynamics, 105, 409,
  \dodoi{10.1080/03091929.2010.509927}

\bibitem[{{Showman}(2007)}]{Showman_2007}
{Showman}, A.~P. 2007, Journal of Atmospheric Sciences, 64, 3132,
  \dodoi{10.1175/JAS4007.1}

\bibitem[{Skamarock \& Klemp(1992)}]{Skamarock&Klemp1992}
Skamarock, W.~C., \& Klemp, J.~B. 1992, Monthly Weather Review, 120, 2109 ,
  \dodoi{10.1175/1520-0493(1992)120<2109:TSOTSN>2.0.CO;2}

\bibitem[{{Skinner} \& {Cho}(2021)}]{Skinner2021}
{Skinner}, J.~W., \& {Cho}, J.~Y.~K. 2021, \mnras,
  \dodoi{10.1093/mnras/stab2809}

\bibitem[{{Tabataba-Vakili} {et~al.}(2020){Tabataba-Vakili}, {Rogers},
  {Eichst{\"a}dt}, {Orton}, {Hansen}, {Momary}, {Sinclair}, {Giles},
  {Caplinger}, {Ravine}, \& {Bolton}}]{Vakili+2020}
{Tabataba-Vakili}, F., {Rogers}, J.~H., {Eichst{\"a}dt}, G., {et~al.} 2020,
  \icarus, 335, 113405, \dodoi{10.1016/j.icarus.2019.113405}

\bibitem[{{Vallis}(2006)}]{VallisBook}
{Vallis}, G.~K. 2006, {Atmospheric and Oceanic Fluid Dynamics},
  \dodoi{10.2277/0521849691}

\bibitem[{{Vallis} {et~al.}(2018){Vallis}, {Colyer}, {Geen}, {Gerber},
  {Jucker}, {Maher}, {Paterson}, {Pietschnig}, {Penn}, \&
  {Thomson}}]{Isca_2018}
{Vallis}, G.~K., {Colyer}, G., {Geen}, R., {et~al.} 2018, Geoscientific Model
  Development, 11, 843, \dodoi{10.5194/gmd-11-843-2018}

\bibitem[{{van der Walt} {et~al.}(2011){van der Walt}, {Colbert}, \&
  {Varoquaux}}]{numpy2011}
{van der Walt}, S., {Colbert}, S.~C., \& {Varoquaux}, G. 2011, Computing in
  Science and Engineering, 13, 22, \dodoi{10.1109/MCSE.2011.37}

\bibitem[{{Williams}(1978)}]{Williams1978}
{Williams}, G.~P. 1978, Journal of Atmospheric Sciences, 35, 1399,
  \dodoi{10.1175/1520-0469(1978)035<1399:PCBROJ>2.0.CO;2}

\bibitem[{{Williams}(1988)}]{Williams1988}
---. 1988, Climate Dynamics, 3, 45, \dodoi{10.1007/BF01080901}

\bibitem[{{Yano} \& {Flierl}(1992)}]{YANO1992}
{Yano}, J.-I., \& {Flierl}, G.~R. 1992, Dynamics of Atmospheres and Oceans, 16,
  439, \dodoi{https://doi.org/10.1016/0377-0265(92)90001-A}

\end{thebibliography}
\bibliographystyle{aasjournal}

%% This command is needed to show the entire author+affiliation list when
%% the collaboration and author truncation commands are used.  It has to
%% go at the end of the manuscript.
%\allauthors

%% Include this line if you are using the \added, \replaced, \deleted
%% commands to see a summary list of all changes at the end of the article.
%\listofchanges

\end{document}